\documentclass[%
	 aip,
	 jap,%
	 amsmath,amssymb,
	 reprint,%
	]{revtex4-1}
	
	\pdfoutput=1
	\usepackage{makecell} 
	\usepackage{booktabs} 
	\usepackage{mdframed}

	\usepackage{multirow} 

	\usepackage{graphicx}
	\usepackage{subfigure} 
	\usepackage{epstopdf}
	\usepackage{dcolumn}
	\usepackage{bm}
	\usepackage{subfigure}
	\usepackage[mathlines]{lineno}
	\def\sam{\textrm{sam}}
	\def\ins{\textrm{ins}}
	\def\skin{\textrm{skin}}
	\def\meas{\textrm{meas}}
	\def\refrm{\textrm{ref}}
	\def\total{\textrm{total}}
	
\begin{document}

        
	\title[Modulation calorimetry in DACs I]{Modulation calorimetry in diamond anvil cells I: heat flow models}
	\author{Zachary M. Geballe}
	\affiliation{Geophysical Laboratory, Carnegie Institution for Science, Washington, DC 20015}
	\author{Gilbert W. Collins}
	\affiliation{Lawrence Livermore National Lab, Livermore, CA 94550}
	\author{Raymond Jeanloz}
	\affiliation{Department of Earth and Planetary Science, University of California, Berkeley, CA 94720}
	\date{\today}	

	\begin{abstract}
Numerical simulations of heat transport in diamond anvil cells reveal a possibility for absolute measurements of specific heat via high-frequency modulation calorimetry. Such experiments could reveal and help characterize temperature-driven phase transitions at high-pressure, such as melting, the glass transition, magnetic and electric orderings, or superconducting transitions. Specifically, we show that calorimetric information of a sample cannot be directly extracted from measurements at frequencies slower than the timescale of conduction to the diamond anvils (10s to 100s of kHz) since the experiment is far from adiabatic. At higher frequencies, laser-heating experiments allow relative calorimetric measurements, where changes in specific heat of the sample are discriminated from changes in other material properties by scanning the heating frequency from $\sim 1$ MHz to 1 GHz. But laser-heating generates large temperature gradients in metal samples, preventing absolute heat capacities to be inferred. High-frequency Joule heating, on the other hand, allows accurate, absolute specific heat measurements if it can be performed at high-enough frequency: assuming a thin layer of KBr insulation, the specific heat of a 5 $\mu$m-thick metal sample heated at 100 kHz, 1 MHz, or 10 MHz frequency would be measured with 30\%, 8\% or 2\% accuracy, respectively. 

\end{abstract}
		
	\newcommand{\ud}{\mathrm{d}}
          \maketitle
 \section{Introduction}
          
Historically, calorimetry experiments at ambient pressure have led to an abundance of new physics, from Einstein's quantum theory of solids \cite{Pais1979} and Debye's vibrational model,\cite{Debye1912} to the discovery of superfluidity \cite{Donnelly1995} and a recent test of renormalization-group theory.\cite{Lipa1996} In Earth science, calorimetric measurements of silicates have proven essential for understanding magmatic processes.\cite{Lange1990}

Some research has aimed to make the same type of measurements at high pressures ($>$ 1 GPa). A few groups have demonstrated the ability to detect phase transitions at low temperatures (tens of mK to tens of K) in heavy fermion compounds up to $\sim 10$ GPa. \cite{Demuer2000,Fernandez-Panella2011,Wilhelm1999,Bouquet2000,Sidorov2010} In making these measurements, they have shown that qualitative calorimetry inside diamond cells is possible, but it is unclear whether such measurements are quantitatively accurate, even in relative values of specific heat as temperature is varied. In some dynamic high-pressure experiments, pyrometry-based temperature measurements allow calculation of specific heats as a shock wave decays.\cite{hicks2006,Eggert2009} However, the short timescales (nanoseconds) of the experiments can result in large uncertainties in the measurement, including the degree and nature of equilibrium achieved, and the large strain rates may generate a high density of atomic defects. Overall, although some pioneering experiments have been carried out, little is known about the specific heat of materials at pressures above a few GPa, or even about the potential for making calorimetric measurements at high pressures. 

If successfully developed, high-pressure calorimetry techniques could have far-reaching applications. They could aid in the detection and characterization of high-pressure entropy-driven transformations, such as melting and order-disorder transitions.\cite{Hazen1996} The energetics of high-pressure phase transitions could reveal new information relevant to Earth science and to materials physics, such as the latent heat of melting in the Earth's core or the pressure-dependence of pre-melting phenomena. At low temperatures, calorimetry could be used to map out first- and second-order phase boundaries in pressure-temperature space, possibly helping in the search for quantum phase transitions (e.g. Ref. \onlinecite{Petrova2012}).

We focus here on calorimetry in the diamond-anvil cell because such cells currently achieve the highest pressures under static compression, and they allow samples to be probed by numerous other techniques (spectroscopy, diffraction, etc.). Static methods offer the widest range of time periods over which the sample can be probed, and may be essential for reaching thermodynamic equilibrium.

Before presenting the main model and results, we explain the need for new modeling of modulation calorimetry, a technique that is over 100 years old. Also, to show the potential for realization of the technique modeled here, we outline available technologies.

\section{Barriers to the use of traditional calorimetry models and methods}

Adaptation of typical calorimetric measurements to high pressures is a challenge because the volume at pressure - thermal insulation as well as sample - is small and therefore difficult to insulate thermally; relevant dimensions are of order $\mu$m to tens of $\mu$m so it is difficult to maintain adiabatic conditions. Moreover, insulating materials are liquids and solids with negligible porosity, hence large thermal conductivity and heat capacity, causing many equations of modulation calorimetry to be invalid.

Most modulation calorimetry experiments are analyzed using an approximate solution to the heat equation, with a single timescale of heat loss. Specifically, the sample is assumed to be thermally connected to a heater via a link with thermal conductance $K_h$, to a thermometer via $K_t$ and to a thermal bath via $K_b$.\cite{Sullivan1968, Kraftmakher2004} In this sense, the sample and its surroundings are modeled as if they are disjoint pieces. Heat flow to the thermal bath is assumed to be small and temperature gradients between the sample, heater and thermometer are assumed to be negligible. 

The result, which is not true in diamond cells, but which provides a baseline to compare against, is that the sample's temperature varies according to
\begin{equation}
T_\omega = \frac{p_\omega}{\sqrt{C^2\omega^2 + K_b^2}}
\label{eqn:basic}
\end{equation}
where $T_\omega$ is the amplitude of temperature oscillation at the given frequency, $\omega$, $p_\omega$ is the amplitude of power oscillation, and $C$ is the heat capacity of sample and addenda. \footnote{This is Eq. (2.3c) of Ref. \onlinecite{Kraftmakher2004} with $Q'$ instead of $K_b$, Eq. (2) of Ref. \onlinecite{Baloga1977} with $\Gamma$ instead of $K_b$, Eq. (1) of Ref. \onlinecite{Bouquet2000}, or Eq. (11) of Ref. \onlinecite{Sullivan1968} with $\tau_1 = C/K_b$, $\tau_2 \to 0$, and $K_b/K_s \to 0$.} 

\begin{mdframed}
Definition: \hspace{5 mm}
The \textit{addenda} are the materials close to the sample that heat diffuses into and away from during a heating cycle, effectively adding an apparent thermal mass to the sample. They include the heater and thermometer if they are separate from the sample, as well as any material that is within $\sim 1$ thermal diffusion length of the sample.
\end{mdframed}
To solve for both heat capacity and thermal conductance to the bath, Eq. (\ref{eqn:basic}) can be fitted to data at variable frequency, or a second equation can be used:
\begin{equation}
\tan\phi = C\omega/K_b
\label{eqn:phase}
\end{equation}
where $\phi$ is the phase shift between heat-source and temperature oscillations at a given frequency. 

In any high-pressure system, including diamond-anvil cells, a model of disjoint sample, heater, thermometer, and thermal bath is not reasonable since the sample is contiguous with other liquids or solids on all sides, resulting in at least two problems: the addenda contribution to measured heat capacity can be large, and the extent of the addenda can depend on frequency; at lower frequencies thermal diffusion extends further into the surrounding material. Indeed, Ref. \onlinecite{Baloga1977} has shown that a 500 $\mu$m-thick Invar sample heated at 0 to 2 Hz in a pressure cell with $\leq 170$ MPa argon gas surrounding the sample is not well described by Eq. (\ref{eqn:basic}). Rather, the addenda contribution increases from approximately zero at ambient pressure (their Fig. 2) to $\sim 100\%$ at 150 to 170 MPa (their Fig. 3). \footnote{Ref. \onlinecite{Baloga1977} uses an analytic expression for an addenda contribution to heat capacity at variable frequency and shows that it fits their data.}

But at high enough frequency, it is possible that nanogram samples inside diamond cells could be heated in a manner that is close enough to adiabatic so that the error in measured value of a sample's heat capacity is small. For comparison, our previous work shows that heating timescales of 1 ns to 1 $\mu$s are required for pulsed heating experiments to reveal the latent heat expected during melting or other first order phase change of 1 $\mu$m-thick metal samples.\cite{Geballe2012}

\section{Available technologies for calorimetry in diamond-cells}

Several existing technologies can be exploited in the design of modulation calorimetry in diamond cells. Examples of specific Joule- and laser-heating designs are described in Appendices A,B.

Joule-heating, which allows heat to be deposited inside metallic samples, can be accomplished by lithographically fabricating wires and metallic samples onto the diamonds or onto a thin layer of thermal insulation,\cite{Weir2009a} or by positioning thin foils between insulating layers and through an electrically-insulating gasket.\cite{Zha2004,Komabayashi2009} Tapered electrical leads can connect the $\mu$m sized sample to electrical connectors that connect to commercial AC power supplies capable of outputting waveforms at kHz, MHz or GHz frequencies. Voltage measuring leads can be connected to lock-in amplifiers or analog-to-digital converters that monitor power oscillations, and perhaps also temperature oscillations via the third harmonic technique described in Appendix C. The background temperature can be measured using a thermocouple if the entire diamond-cell is heated, or by spectroradiometry if the sample's temperature is at least 1000 K. A more detailed Joule-heating design is presented in Part II of this publication.

Laser-heating, which deposits heat on the surface of metallic samples, can be accomplished with 100 MHz frequency oscillations with commercial diode laser modules (e.g. Newport LQD series), or with several GHz using more-specialized electrical modulation of a diode laser source (e.g. Ref. \onlinecite{Melentiev2001}). Pyrometric or spectroradiometric temperature measurements can be made with fast light-collecting technologies, such as intensified CCD cameras and photodiodes with nanosecond resolution. Incident laser power can be measured at the laser source by using a power meter, while changes in laser absorption can be monitored by measuring reflectivity from the sample area using a photodiode.

\begin{table}
\begin{center}
\begin{tabular}{ {c} p{1.3 cm} {c} {c}} 
& & \multicolumn{2}{c}{Temperature measurement} \\
\cline{3-4}
& \multicolumn{1}{c|}{} & Internal & \multicolumn{1}{c|}{Surface}\\
\cline{2-4}
Heat& \multicolumn{1}{|c|}{Internal} & I/I & \multicolumn{1}{c|}{I/S} \\
Source & \multicolumn{1}{|c|}{Surface} & - & \multicolumn{1}{c|}{S/S} \\
\cline{2-4}
\end{tabular}

\caption{Summary of possible calorimetry designs with two-letter labels for the combinations studied here.}
\label{table:summary}
\end{center}
\end{table}

\section{Scope of analysis}
Our analysis discriminates between three types of temperature measurement and heating scheme, summarized in Table \ref{table:summary}: internal heating experiments with internal temperature measurement (``I/I'', such as Joule heating with the third harmonic temperature measurement technique), internal heating experiments with surface temperature measurement (``I/S'', such as Joule heating with spectroradiometry), and surface heating experiments with surface temperature measurement (``S/S'', such as laser heating of metals with pyrometry or spectroradiometry). We do not consider ``S/I'' because most experimental designs that allow for internal temperature measurement also allow for an internal heating source (I/I), which is likely to give more accurate calorimetry results because no heat transport is required to equilibrate the heated region with the region of temperature measurement. For example, if the amplitude of temperature oscillations is determined by measuring electrical resistance, then resistive (Joule) heating could also be used to deposit heat; if temperature is determined from electromagnetic radiation that is interior to the sample, it is likely that electromagnetic radiation (e.g., time-modulated laser heating) could also be deposited in the sample's interior. More nearly adiabatic conditions result from using the internal heating source in both cases.

We do not consider heating sources or temperature measurements that are far away from the sample because this increases the difficulty in extracting calorimetric information about the sample itself. To give a sense of the difficulty, typical diamond anvils are $\sim 2$ mm in each linear dimension, giving a thermal mass that is $\sim 10^5$ times the thermal mass of a typical diamond cell sample (10 $\mu$m thick, 100 $\mu$m in diameter). The gasket and epoxy that border the diamonds are also large compared to the sample, and their heat capacities would be difficult to calibrate since their dimensions typically vary from experiment to experiment.

We do not explicitly consider heating of samples by the ``hot plate'' method, in which one material (usually a thin metal foil) absorbs heat, which then diffuses into the sample of interest, despite the fact that this method can be useful for measuring a variety of properties (e.g. inteface conductance\cite{Cahill2004} and thermal diffusivity \cite{Cahill1990,Beck2007,Imada2014}). In fact, it is the diversity of uses that makes this heating method difficult to analyze comprehensively; temperature evolution is affected by both transport and thermodynamic properties of both the heat absorber and the sample. Nonetheless, we expect that thermal measurements using hot plate heating in diamond cells will be useful in the future, and therefore discuss the topic briefly in the discussion sections of this paper and the companion paper.

For simplicity, we assume local thermodynamic equilibrium is reached within the heated area, ignoring the possibility of kinetic barriers to phase transitions even though kinetics are known to affect many phase transformations. \cite{Birge1997}  One way to account for, or at least estimate the influence of, kinetic effects is to reverse transformations by slowly raising and then slowly lowering temperature.

 In principle, measurement of phase shifts could be used in conjunction with the amplitude of oscillations and two equations, (\ref{eqn:basic}) and (\ref{eqn:phase}), to solve for the heat capacity and thermal conductance of the link to the temperature bath. Unfortunately, Eq. (\ref{eqn:phase}) is not a good approximation for diamond-cell samples, resulting in little improvement at frequencies greater than 1 MHz (Appendix D).

\begin{table*}
\begin{center}
\begin{tabular}{ p{3 in} p{1.5 in} c } 
& Sample (Fe) & Insulator (KBr)\\
\hline 
$d$: Layer thickness ($\mu$m) & 5 & 10 \\
$\rho$: Density (g cm$^{-3}$) & 7.9 & 2.75 \\
$c$: Specific heat (J g$^{-1}$ K$^{-1}$) & 0.45 & 0.45 \\
$k$: Thermal conductivity (W m$^{-1}$ K$^{-1}$) & 80 & 4.8 \\
$D$: Thermal diffusivity ($\mu$m$^2$ $\mu$s$^{-1}$) & 22 & 3.9 \\
$r$: Resistivity ($\Omega$ m) & $9.7 \times 10^{-8}$ & - \\
$d\textrm{log}r/dT$: Temperature coefficient of resistance (K$^{-1}$) & 0.0064 & - \\
$\delta_{\skin}$: Skin depth ($\mu$m)& 
1000 if internal & - \\
& 0.1 if surface & \\

\hline

\end{tabular}

\caption{Properties of the sample and insulator used in our reference simulations.}
\label{table:mat_props}
\end{center}
\end{table*}

\section{Modeling scheme}
We model heat flow during modulated heating of metal samples pressed between symmetric layers of thermal insulation in a diamond anvil cell, as depicted schematically in Figs. \ref{fig:RH_schematic}, \ref{fig:LH_schematic}. Metals are chosen in this study since they are easier to heat via Joule-heating or laser-heating, and because they are likely choices for use as standard heaters in the future. We typically assume a total thickness from diamond to diamond of 30 $\mu$m, with sample thickness ranging from 1.7 to 15 $\mu$m, the remaining space consisting of thermal insulation between diamond and sample. We approximate the heat flow in the central part of the heated area as occurring solely in the axial direction of the diamond anvil cell (``$z$'' in Fig. \ref{fig:LH_schematic}), allowing reduction of the heat equation to one dimension along the axis:
\begin{equation}
\frac{\partial T(z,t)}{\partial t} = \frac{1}{\rho C} \frac{\partial (k\frac{\partial T}{\partial z})}{\partial z} + Q(z,t)
\label{eqn:T_diff}
\end{equation}
where $T$ is temperature at time $t$ and position $z$, $Q$ is a heating source, and $\rho$, $C$, and $k$ are material properties defined in Table \ref{table:mat_props} (assumed to be temperature-independent). In a previous publication, we used examples of two-dimensional axial simulations to show that the assumption of purely axial heat flow in diamond cells is valid at high frequencies ($f > D_{sam}/\textrm{width}$).\cite{Geballe2012} 

We assume two heating sources of equal power are distributed through a skin depth, $\delta_{skin}$, from each side and that they vary sinusoidally in time with frequency $f$:
\begin{equation} 
Q(z,t) = p_0\left( e^{-\frac{z-d_{\sam}/2}{\delta_{\skin}}} +e^{-\frac{z+d_{\sam}/2}{\delta_{\skin}}}\right)\sin(2\pi f t) 
\label{eqn:Q}
\end{equation}
for $-d_{\sam}/2 < z < d_{\sam}/2$ and $Q(z,t) = 0$ otherwise. In Appendix E, a different heating source is used to simulate one-sided heating. We also tested a more realistic heating source that is always positive, $Q \propto 1+\sin(2\pi f t)$, and found no change in the resulting temperature variations.\footnote{The only change due to addition of a constant background heating source to a sinusoid (e.g. $Q \propto 1+\sin(2\pi f t)$ or $Q \propto 100 + \sin(2\pi f t)$) is in time-averaged temperature profile. The dynamic temperature response is identical.}

Layer thicknesses and material properties used in our reference simulations are listed in Table \ref{table:mat_props}. The material properties match ambient pressure-temperature values of an iron sample (except with a relative magnetic permeability of 1) and a single crystal potassium bromide insulator.

As in Refs. \onlinecite{Geballe2012,kiefer2005}, we assume that thermal conduction through the diamond anvils is so efficient that the temperature at the culet surface (diamond tip) is a constant (e.g. 300 K). In the main text, symmetric heating allows us to simulate one half of the sample chamber, from $z =0$ to $z= d_\sam + d_\ins$, as long as we enforce the boundary condition that no heat flows across the mid-plane of the sample, $\frac{\partial T}{\partial z}|_{z=0} = 0$. 
The initial condition is that $T$ is constant in space, which is the average temperature distribution assuming the heating source described in Eq. (\ref{eqn:Q}).

We solve Eq. (\ref{eqn:T_diff}) by implementing the Crank-Nicholson numerical method described in Appendix F, with a typical time step of $0.01/f$ for frequency $f$, and a 10 to 50 nm mesh (i.e. 100-times smaller than the smallest sample dimension). We typically simulate 10 heating cycles and fit the final 5 cycles to a sinusoidal function in order to extract an amplitude of temperature oscillation. For laser-heating, we assume that the measured temperature is a weighted average of temperature to the fourth power in order to approximate the effect of the Stefan-Boltmann law:
\[T_{\meas} = \left(A \int_0^{d_{\sam}}{T^4(z) e^{-\left(\frac{z-d_{\sam}}{\delta_{\skin}}\right)}dz}\right) ^{1/4} 
\]
where A is a normalization factor. In the case of Joule heating with third harmonic temperature measurement described below, we average $T$ rather than $T^4$, and since the skin depth is large compared to sample thickness, the exponential term approaches $e^0$. Hence,
\[T_{\meas} = A \int_0^{d_{\sam}}{T(z)dz} \]

The total heat capacity of sample plus addenda that would be inferred, is
\begin{equation}
C_\total = \frac{p_\omega}{\omega T_\omega}
\label{eqn:C_meas_simple}
\end{equation}
where $p_\omega$ is the amplitude of power oscillations that is absorbed through the full thickness of the sample ($p = \int_{-d_{sam}}^{d_{sam}}{Q(z,T)dz}$), and $T_\omega$ is the amplitude of oscillation of $T_{\meas}$, and $\omega$ is the angular frequency of both oscillations. \footnote{In the case of Joule-heating, we will redefine $\omega$ as the angular frequency of current or voltage oscillations, which are 2-fold smaller than the frequency of power oscillations, meaning ``$\omega$''s in Eq. (\ref{eqn:C_meas_simple}) will be replaced by ``$2\omega$''s in this paper and in Part II.}

\begin{mdframed}
Note: \hspace{5 mm} The variable $c$ or $C$ is used in several contexts in this study. Lower-case $c_{\sam}$ and $c_{\ins}$ are specific heats of sample and insulation material (units: Jg$^{-1}$K$^{-1}$, i.e. the material property).

Upper-case $C$ is the apparent heat capacity given by the ratio of energy input divided by temperature change, whether or not the conditions are adiabatic: in gneray they are not (units: J/K, i.e., the product of specific heat and mass). A subscript or superscript ``total'' implies the total heat capacity of sample plus addenda that would be inferred from power deposited and temperature oscillation measured, as opposed to a true heat capacity of sample or insulator. Superscript ``ref'' refers to the reference properties of Table \ref{table:mat_props}. For example, $C_\total^\refrm$ is the total heat capacity that would be measured according to the simulation results using the reference properties.

\end{mdframed}

\begin{figure}
\begin{center}
\includegraphics[width=3.5in]{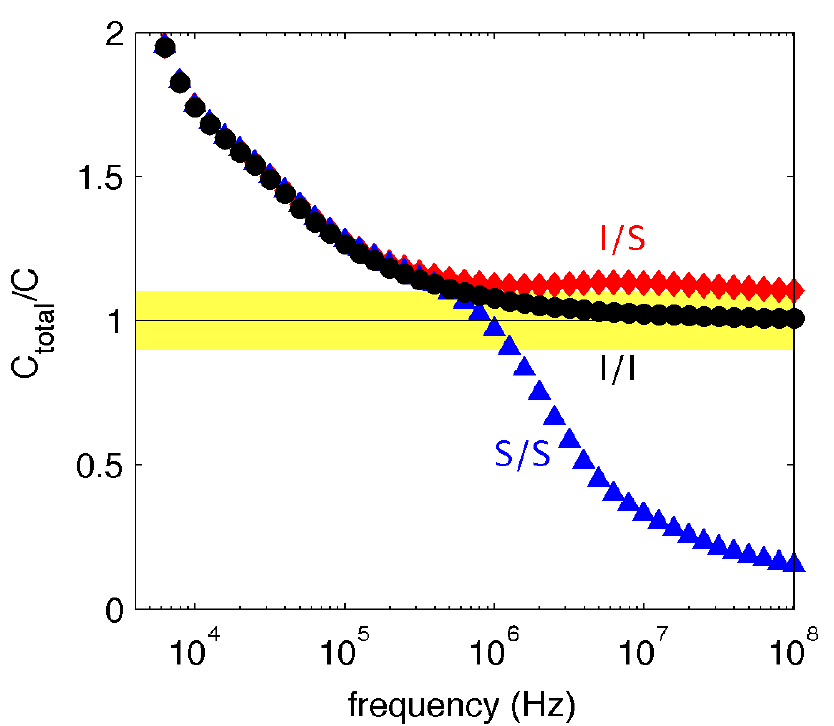} 
\caption{Total heat heat capacities per unit area ($\frac{p_\omega}{\omega  T_\omega}$) divided by sample heat capacity per unit area ($\rho_{\sam}c_{\sam}d_{\sam}$) as a function of the frequency of heating modulation. Black circles indicate that heat is deposited internally and temperature is measured internally in our reference experiment, red diamonds indicate that heat is deposited internally and temperature is measured at the surface ($\delta_{\skin} = 100$ nm), and blue triangles indicate that both temperature measurement and heat deposition take place in the surface ($\delta_{\skin} = 100$ nm). The yellow band marks $< 10 \%$ error in heat capacity measurement.}
\label{fig:int_surf_comp}
\end{center}
\end{figure}

\begin{figure}
\begin{center}
\includegraphics[width=3.5in]{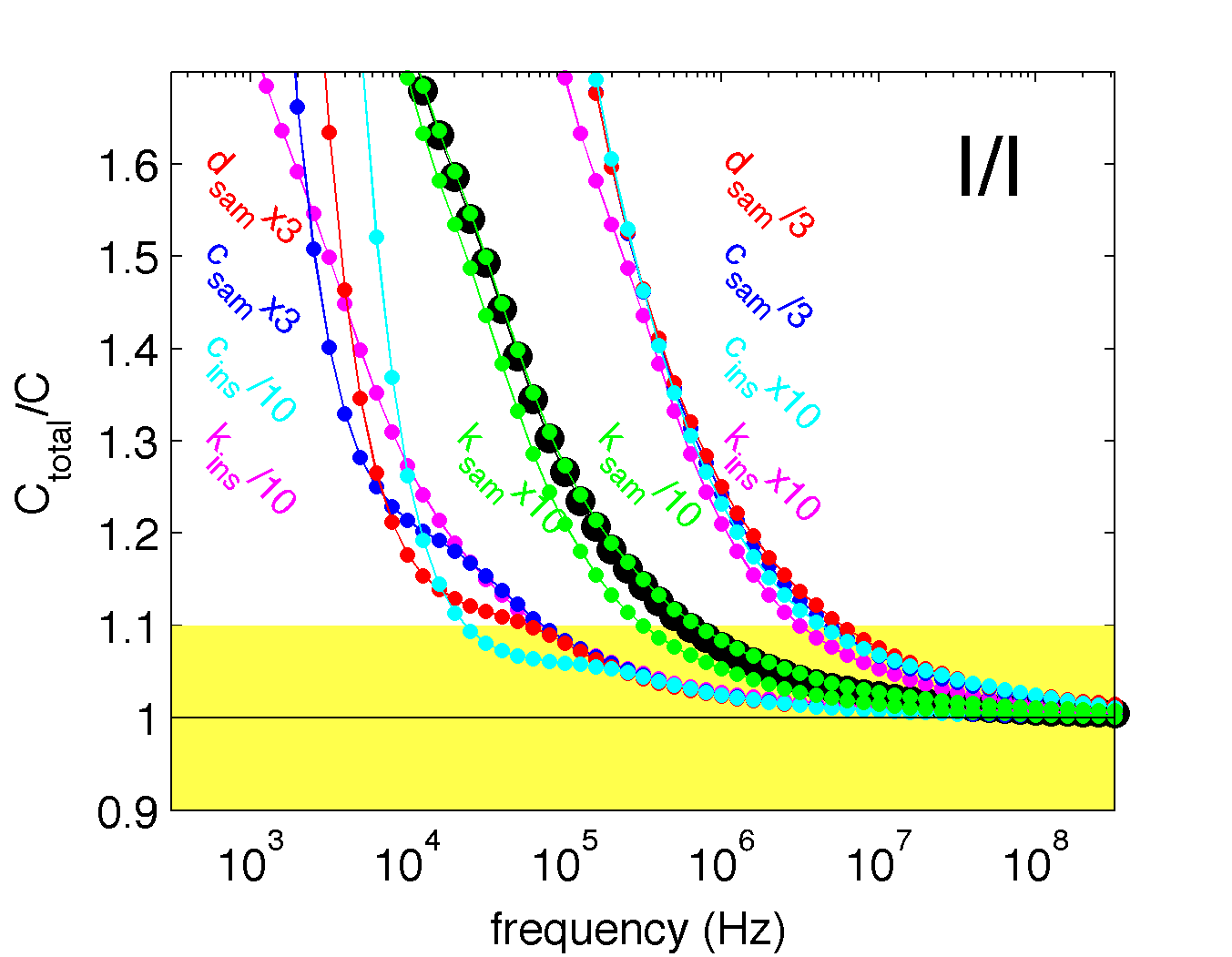} 
\caption{Error in heat capacity of sample due to addendum contribution vs. heating frequency. Black circles represent the reference experiment with internal heating and temperature measurement, while colors indicate that a single parameter has been changed by a factor of 10 from the reference. Yellow highlights the region with $< 10\%$ error.}
\label{fig:cal_errs_var_props}
\end{center}
\end{figure}

\begin{figure*}
\begin{center}
\includegraphics[width=5in]{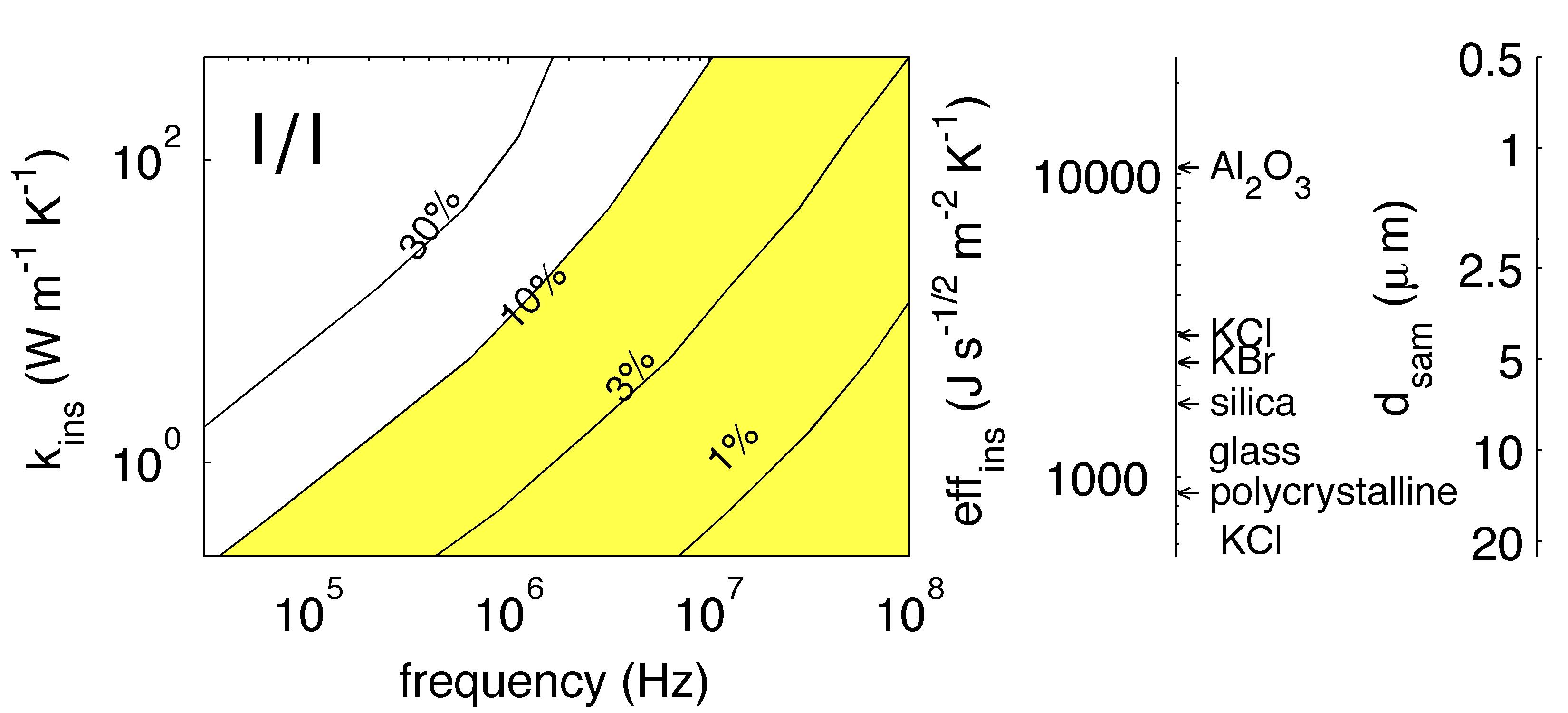}
\caption{Contours of error in measurement of heat capacity due to addendum contribution in a parameter space that describes two key properties of the proposed Joule-heating experiments: thermal conductivity of the insulation and the frequency of heating. Yellow shading marks  $< 10\%$ error. We assume internal heating, internal temperature measurement, reference material properties (except for thermal conductivity, which varies), and reference geometries. The y-axes to the right of the figure show alternative changes from the reference experiment (KBr insulation, 5 $\mu$m-thick sample) that would result in approximately the same addenda contribution as the thermal conductivity of insulation plotted on the left-hand-side. In particular, we have mapped the effected of changes in thermal conductivity of insulation, $k_\ins$, onto changes in thermal effusivity, $\sqrt{\rho_\ins c_\ins k_\ins}$ or changes in sample thickness.}
\label{fig:K_f_limits}
\end{center}
\end{figure*}

\begin{figure*}
\begin{center}
\includegraphics[width=6.5in]{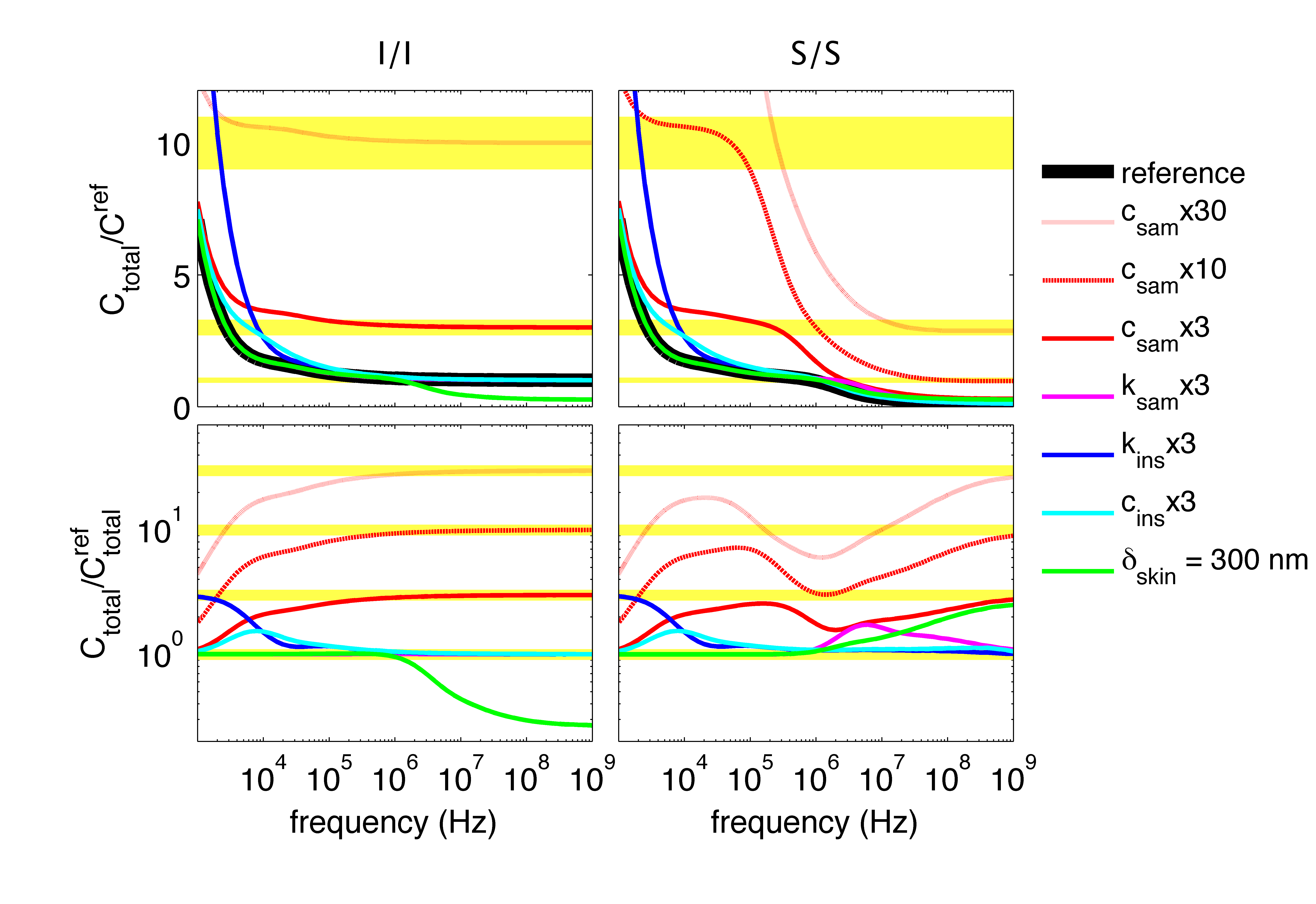} 
\caption{Total heat capacity of sample plus addenda (top row) and ratios of total heat capacity to total heat capacity in the reference model (bottom row) upon variation of the sample's specific heat (red curves), the sample's thermal conductivity (pink curves), the insulation's thermal conductivity (blue curves), the insulation's specific heat (cyan curves), and the skin depth of the heating source and temperature measurement probe (green curves). All four horizontal axes show frequency. The vertical axes in the top row are total heat capacities normalized by the true value of heat capacity of the sample. The vertical axes in the bottom row are the same total heat capacities, but normalized by the total heat capacity of sample plus addenda in the reference state ($C_\total^{\refrm}$). The first column assumes internal heating and temperature measurement (I/I), while the second column assumes surface heating and temperature measurement (S/S, for which $\delta_{\skin} = 100$ nm). Shades of red indicate the magnitude of specific heat changes (see legend), while yellow highlights accurate measurements (within 10$\%$) for 1-fold, 3-fold, 10-fold and 30-fold increases in sample heat capacity.} 
\label{fig:C_changes}
\end{center}
\end{figure*}

\section{Results}
First, we assume the material properties and sample-chamber dimensions listed in \ref{table:mat_props}. We simulate heating experiments at 4 kHz to 100 MHz using the three geometries of heat source/temperature measurement described above, (I/I), (I/S) and (S/S). The total heat capacity inferred from power and temperature oscillations is shown in Fig. \ref{fig:int_surf_comp}. 

At frequencies low compared to the timescale of heat conduction out of the sample, total heat capacity includes a large addenda contribution. For example, at 9 kHz, the addenda contribution is as large as the sample contribution to heat capacity (i.e. $C_\total/C = 2$).

Near 1 MHz frequency, the total heat capacity drops to within $\sim 10\%$ of the sample's heat capacity in all cases (i.e. $1.1 > C_\total/C > 0.9$) . At higher frequencies, the proposed measurement becomes even more accurate in the case of (I/I): total heat capacity asymptotically approaches the sample's value, while the addenda contribution becomes negligible. Examples showing this approach to adiabatic heating are shown in Appendix A, along with an idea of how it could be experimentally realized.

In the other cases, we also expect addenda contribution to heat capacity to become small at high frequency, but other details cause complications. In the case of I/S, the temperature measurement is near the sample surface, which is highly susceptible to thermal diffusion into the insulation. In fact, if temperature were measured at the true interface between sample and insulation, we expect insulation to contribute a measurable addendum at all frequencies. Here, we have assumed a 100 nm skin-depth, causing the slow decrease of total heat capacity at 10 to 100 MHz, when the lengthscale of diffusion starts to approach the skin-depth. 

High-frequency experiments using (S/S) involve further complication. In this case, heat diffuses to all material near the sample-insulation interface (i.e. into both sample and addenda), but not to the sample's interior (see Appendix B for an example). Hence, at high frequency, total heat capacity is far smaller than the full sample's heat capacity. There is one lucky frequency at which addenda additions exactly cancel the reductions in heated sample, but its value depends on several material properties, including two that cannot be measured easily in a diamond cell: the skin depths of the heating source and of the temperature measurement probe.

To gain more insight into the total heat capacities that would be measured in Joule-heating with internal temperature measurement (I/I), we simulate a range of other experiments using the (I/I) scheme. A 3-fold increase in the sample's total heat capacity (i.e. a denser, greater specific-heat, or thicker sample) causes a $\sim 10$-fold reduction in the frequency requirement for low-addenda heat capacity measurements (Fig. \ref{fig:cal_errs_var_props}) The same effect results from 10-fold decreases in density, specific heat, or thermal conductivity of insulation. Thermal conductivity of the sample has a relatively small effect.

To state our findings more succinctly, we introduce the term ``thermal effusivity'', a material property that describes the ability of a material to absorb heat from its surface via conduction, which is defined by $\sqrt{\rho c k}$, with $\rho$, $c$ and $k$ being density, heat capacity and thermal conductivity. We can summarize the previous two observations by the following: a 3-fold decrease in the ratio of thermal effusivity of insulation to total heat capacity of sample causes a 10-fold decrease in frequency needed to achieve a fixed value of addenda contribution to total heat capacity. The inverse is also true: a 3-fold decrease in insulation thermal effusivity to sample heat capacity causes a 10-fold increase in frequency requirement. 

Motivated by the large effect of insulation effusivity, a property that can be tuned over a wide range, we extend our calculations to more extreme values. In particular, we vary the thermal conductivity of insulation by two orders of magnitude in each direction, and present the results as contours of heat capacity measurement error in Fig. \ref{fig:K_f_limits} . Also, we project these simulated errors onto two other axes, the effusivity of the insulation and the thickness of metal sample, where we assume the addenda contribution is a function of frequency and the single variable $\sqrt{\rho_\ins c_\ins k_\ins}/(\rho_\sam c_\sam d_\sam)$, as suggested by Fig. \ref{fig:cal_errs_var_props}. The result, Fig. \ref{fig:K_f_limits}, is  meant as a guide for design of experiments. For example, use of polycrystalline KCl \cite{El-Sharkawy1984} or silica glass instead of single crystal KBr improves accuracy of heat capacity measured, whereas use of single crystal alumina decreases accuracy. Motivated by this observation, we chose to use silica glass instead of KCl in Part II of this two-part publication.

Variations in sample thickness also cause a large effect in the accuracy of measured heat capacity, with 20 $\mu$m-thick samples allowing frequencies less than 100 kHz to result in heat capacity measurements with 10$\%$ accuracy. We also note that a possible trick to reduce error in heat capacity measurement is to measure heat capacities at two (or more) sample thicknesses. The thinner sample is more sensitive to addenda than the thicker one, allowing deconvolution of the two contributions to measured heat capacity: the sample's contribution and the addenda's contribution. Detailed model results are presented in Appendix G.

Finally, in order to guide experiments seeking to detect changes in material properties (rather than absolute values), we study the frequency-dependence of changes in total heat capacity as various material properties change. Fig. \ref{fig:C_changes} shows the results for I/I and S/S schemes. The I/S scheme has been omitted as the results are similar to the S/S results. We normalize the results in two ways, via the heat capacity of the sample alone before changing the material property, $C^\refrm$, and via the heat capacity of sample plus addenda before changing the material property, $C^\refrm_\total$. The former normalization may be more intuitive, but only the latter normalization would be experimentally feasible, so we discuss it here. At $\geq 300$ kHz and $\geq 1$ GHz, the ratio of total heat capacities, $C_\total/C^\refrm_\total$, approaches the correct values for (I/I) and (S/S), respectively. At lower frequencies (10 kHz and 10 MHz), measured heat capacities are more sensitive to increases in heat capacity than to all other changes in material properties considered here ($k_{\ins}$, $c_{\ins}$, $k_{\sam}$), except for a change in skin depth, $\delta_{\skin}$. Therefore, by scanning frequency by a few orders of magnitude around 10 kHz (for Joule-heating, I/I) or 10 MHz (for laser-heating, S/S), increases in heat-capacity may be identified with little ambiguity.

\section{Discussion}
Among the possible experiments considered here, two types of calorimetry measurements are shown to be feasible: absolute measurements of the heat capacities of Joule-heated metals with internal temperature measurement (I/I) and relative measurements of Joule- or laser-heated metals with surface temperature measurement (I/S) or (S/S). In both cases, high heating frequencies must be used ($\sim 100$ kHz to 10 MHz), and to allow for the most robust interpretations, frequency should be varied over such a range.

The requirement of high heating frequency can be understood by analyzing heat flow in the small volume of a diamond cell sample chamber; at frequencies lower than the frequency at which heat leaves the sample, the addenda contribution to heat capacity is large. Specifically, the characteristic timescale for heat conduction through the insulation to the diamonds is $\frac{d_\ins^2}{D_\ins}$, which is 25 $\mu$s for the 10 $\mu$m-thick KBr insulation assumed here. But there is a second source of heat loss: diffusion of heat into the insulation itself. The timescale of heat loss to the insulation is,
\begin{eqnarray}
\tau_{\textrm{into ins.}} 
&=& \left( \frac{\rho_\sam c_\sam d_\sam}{\textrm{eff}_\ins}\right)^2
 \label{eqn:into_ins_timescale}
 \end{eqnarray}
 where $\textrm{eff}_{ins} = \sqrt{\rho_{ins}c_{ins}k_{ins}}$ is the effusivity of the insulation. For the iron sample and KBr insulation assumed here, this timescale is 50 $\mu$s, meaning the insulation becomes a significant part of the addenda at frequencies less than a few tens of kHz.
 
In fact, the insulation contribution to total heat capacity is more difficult to correct for than heat flow to the diamonds anvils. Appendix D shows that total heat capacity can be corrected for contributions from the heat bath (i.e. the diamonds) by measuring the phase shift of the modulated temperature relative to the modulated heating source. Unfortunately, this correction does not seem to account for heat lost to insulation, leaving a large addenda contribution to total heat capacity.

Despite these challenges, we have shown that absolute calorimetry measurements of metals can be made with better than $10\%$ error using Joule heating, and relative heat capacities can be measured using either Joule- or laser-heating. In all cases, careful experimental design (e.g. frequency scanning) is required to discriminate changes in sample specific heat from changes in other properties of the sample or insulation. The simulation results presented here provide a roadmap for such design.

Joule heating affords a controlled way to deposit heat, and the third-harmonic temperature measurement described in Appendix C provides a way to measure temperature oscillations in the exact place that heat is deposited. Key experimental challenges are (1) to build a circuit capable of delivering high-frequency electrical power to the tips of diamonds with little electrical distortion and (2) to press a metal foil of uniform cross-sectional area between well-insulated diamond tips. If such engineering challenges are met, realization of the internal heating/internal temperature measurement experiment modeled here would provide the first absolute measurements of specific heat at $> 10$ GPa of pressure.

Laser heating allows simpler sample preparation and avoids the danger that a broken electrical lead ends an experiment prematurely. Assuming reliable temperature and power measurements, relative heat capacity can be measured at a range of frequencies from $\sim 1$ MHz to 1 GHz, and changes in specific heat can be discriminated from changes in other material properties by referring to Fig. \ref{fig:C_changes} (lower right panel). Indeed, temperature oscillations can be measured via thermal emissions, but estimation of the absorbed fraction of incident laser power is complicated by the potential for changes in reflectivity and opacity of sample and insulation, for example at a phase transition of interest. That is, the possibility of transition-induced changes in material properties would complicate interpretation of experimental measurements, especially if time dependencies (e.g., kinetics) play a role in the frequency range being investigated. Total reflectance and absorbance could be monitored, but even so, there is a possibility that the location of absorbed laser power remains unknown. If the insulation begins absorbing laser power at high temperatures, the model results presented here are of little use. On the other hand, there is little chance (in the absence of an insulator-metal transition) that the electrical conductivity of an insulator increases so much that it significantly alters the location at which Joule-heating power is deposited.

In order to quantify specific heat, the geometry of the sample must be known. The starting material in the proposed Joule-heating experiment should therefore be a strip of metal with uniform cross sectional area, and a soft pressure medium should be used on at least one side of the metal sample so that deformation is limited. Still, shearing is inevitable in diamond-cell experiments, and it typically results in thinning of a sample upon compression. Therefore, an accurate measurement of thickness is crucial to quantify specific heat. At least three options exist: (1) the metal's thickness can be inferred via white-light interference measurements from (i) diamond-to-diamond and (ii) diamond-to-metal on each side, (2) the metal's surface area can be measured \textit{in situ} via optical imaging, or (3) purely elastic strain following a known equation of state can be assumed upon decompression, and the metal's geometry can be measured (or inferred) after decompression to ambient pressure (i.e. \textit{ex situ}). One way to infer sample geometry precisely after decompression is to use the resistance or heat capacity measurement itself and the known ambient pressure value of resistivity or specific heat, as in the high-pressure resistivity measurements of Ref. \onlinecite{Gomi2013}.

A variety of potential benefits of the proposed measurements exist. First, high-pressure phase transitions of metals could be mapped in pressure-temperature space using either Joule- or laser-heating, and phase boundaries that are already documented using structural probes can be confirmed using an entropy-sensitive probe. Little accuracy is required to identify first-order phase transitions; relative heat capacity can be relied upon, and in principle, measured values of heat capacity can differ from true values by a factor of up to $\frac{L}{c_{sam}\sigma_T}$, where $L$ is the latent heat of phase transition and $\sigma_T$ is the precision of the temperature measurement. In fact, several studies of heavy-fermion compounds at high pressures have already documented phase transitions via \textit{relative} specific heat measurements up to $\sim 10$ GPa at $\leq 20$ K.\cite{Demuer2000,Wilhelm1999,Bouquet2000}

Second, the energetics of single phases can be studied and compared against models such as the Debye model. Deviations from models would be especially interesting for the metals that exhibit large increases in specific heat at temperatures 100s of K below melting at ambient pressure (i.e. ``pre-melting'').\cite{Ubbelohde1978}

Once a metal such as tungsten has been well-characterized using high-pressure calorimetry, it can be used as a standard ``hot plate'' heater. The insulator can be replaced with a non-metallic sample of interest, and relative calorimetry experiments can be performed to determine changes in their effusivities (referring to the curves labeled $c_{\ins}$ and $k_{\ins}$ in Fig. \ref{fig:C_changes}). This would complement current techniques that infer thermal diffusivities from pulsed heating experiments, but which require assumptions of heat capacities to infer thermal conductivities. \cite{Cahill2004,Cahill1990,Beck2007,Imada2014} A ``hot-plate'' technique could allow characterization of the energetics of transitions such as substitutional disordering in multi-valent minerals, or the dissociation of molecular fluids. As in the case of mapping pressures and temperatures of temperature-driven transitions in Joule-heated metals, rough calorimetric measurements are likely sufficient to identify the latent heat of many first-order phase transitions, to detect large increases in specific heat near second-order phase transitions, or to discriminate between the two.

We have also described the limitations of calorimetric measurements in diamond cells. First, absolute calorimetry is not possible in surface-heated samples. This precludes the possibility of using laser heating to study the heat capacity of metallic samples (without use of a heat capacity standard). Second, even when using internal heating, the addenda from relatively good thermal insulators such as KBr cause large biases in measured heat capacities at frequencies below $\sim 100$ kHz.

\section{Conclusions}
Many opportunities exist for novel calorimetry experiments at high-pressures using high-frequencies heating sources. Since ``insulators'' in diamond cells are thin layers of dense solids or liquids, significant addenda contributions to measured heat capacities are unavoidable at low frequencies. But those contributions can be limited to $< 10 \%$ of the sample's heat capacity by use of high-frequencies ($\sim 100$ kHz to 10 MHz), internal heating (e.g. Joule heating), relatively low thermal conductivity insulators (e.g. alkali-halides or glasses), and relatively thick samples ($\geq 5$ to 20 $\mu$m).

\appendix

\begin{figure*}
\begin{center}
\includegraphics[width=6.5in]{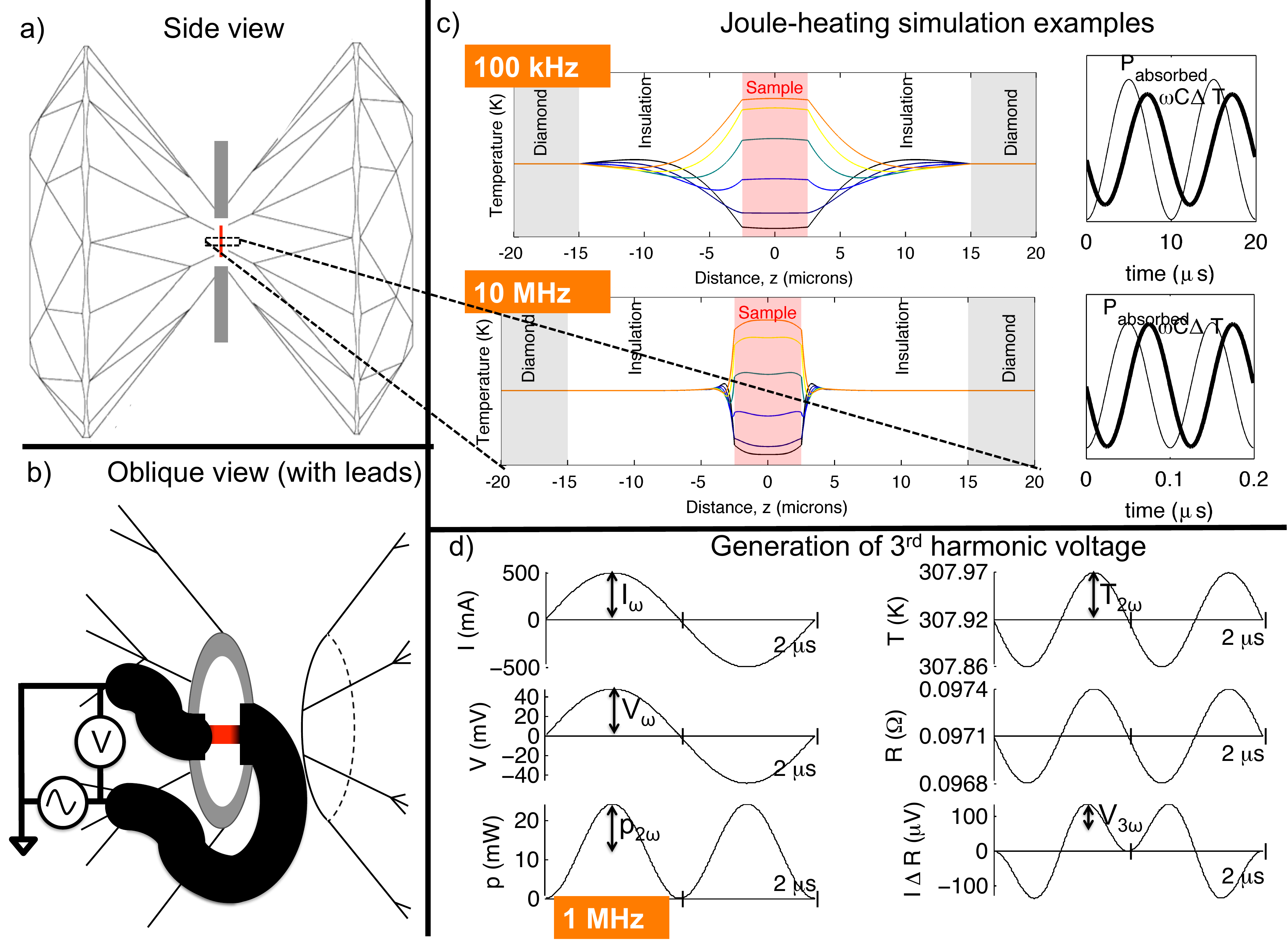} 
\caption{(I/I) Experimental design for absolute measurement of heat capacity at high pressures. (a) Side view of a 5 $\mu$m-thick metal foil (red) surrounded by a transparent thermal insulator (dielectric, shown in white) and compressed in a diamond anvil cell. (b) An oblique view onto the tip of one diamond shows thick electrical leads delivering current to the metal foil. (c) Two examples Joule heating simulations (100 kHz and 10 MHz) showing temperature profiles at various times during half a heating cycle (black to orange represent early to late times). Also shown are the Joule power absorbed and temperature oscillation (times frequency and heat capacity) during two heating cycles. The temperature oscillation curve approaches the power curve in magnitude as the simulation becomes nearly adiabatic at 10 MHz. (d) An example of the dynamic Joule heating and temperature measurement proposed, assuming the geometry and material properties in Table 1. From the top panel, $\pm 500$ mA of 0.5 MHz current is delivered to the metal sample, causing $\pm 40$ mV voltage oscillations at 0.5 MHz, resulting in $\pm 15$ mW power oscillations at 1 MHz on top of a $\pm 15$ mW background power. The temperature rises from 300 K to 308 K due to the background power and oscillates by $\pm$0.05 K, causing $\pm$0.0001 $\Omega$ oscillations in resistance, which feed back into the voltage. The final panel shows this voltage feedback is composed of a first harmonic and a third harmonic. Measurement of the magnitudes of the third harmonics, $V_{3\omega}$, allows calculation of heat capacity by Eq. (\ref{eqn:C_meas}).}
\label{fig:RH_schematic}
\end{center}
\end{figure*}

\section{}
The main text shows that in theory, internal resistive (Joule) heating at $\sim 100$ kHz to 10 MHz frequency can be used to generate quantitative, absolute calorimetric data with limited contamination from addenda. Here we illustrate the principle of such an experiment via a few examples. Fig. \ref{fig:RH_schematic} outlines an experimental setup, and using the numerical simulation described in the main text, shows where heat deposited in the sample would flow during experiments.  A 5 $\mu$m-thick metal sample surrounded by 10 $\mu$m of insulation on each side is compressed between the tips of diamond anvils. Electrical power is deposited in the narrowest part of the circuit (assumed to be the metal sample in the high-pressure chamber) via Joule heating. A small amount of heat flows from the sample into the surrounding insulation at 100 kHz and even less at 10 MHz heating frequency (Fig. \ref{fig:RH_schematic}b).

Heat capacity of samples plus addenda can then be measured electrically: the amplitude of power oscillations is $\frac{1}{2}I_\omega^2R_{\sam}$, and the amplitude of temperature oscillations, $T_{2\omega}$, can be inferred via the third harmonic technique described in Appendix C. Then, assuming a current source drives electricity through the sample and that the temperature-dependence of resistance is known or measured, the total heat capacity of sample plus addenda is
\begin{equation}
C_\total = \frac{\frac{1}{2}(I_\omega)^2R_{\sam}}{2\omega T_{2\omega}} = \frac{(I_\omega)^3(R_{\sam})^2 d\textrm{log}R/dT}{8\omega V_{3\omega}}
\label{eqn:C_meas}
\end{equation}
We have used Eq. (\ref{eqn:C_meas_simple}) with $2\omega$ substituted for $\omega$ (because power oscillates at the second harmonic of current or voltage), and Eq. (\ref{eqn:T2w_part1}) to substitute for $T_{2\omega}$. Part II presents an analogous equation for the more realistic case of a voltage source with its own internal resistance.

Using Eq.  (\ref{eqn:T2w_part1}), a 0.055 K temperature oscillation would be inferred from the 25 $\mu$V-amplitude third harmonic measurement induced by the 1 MHz heating example of Fig. \ref{fig:RH_schematic}d. Using Eq. (\ref{eqn:C_meas}), a total heat capacity of 35.4 nJ/K would be measured, which is a mere 3\% larger than the heat capacity of the $5 \times 20 \times 100$ $\mu$m-sized iron sample.

\begin{figure*}
\includegraphics[width=6.5in]{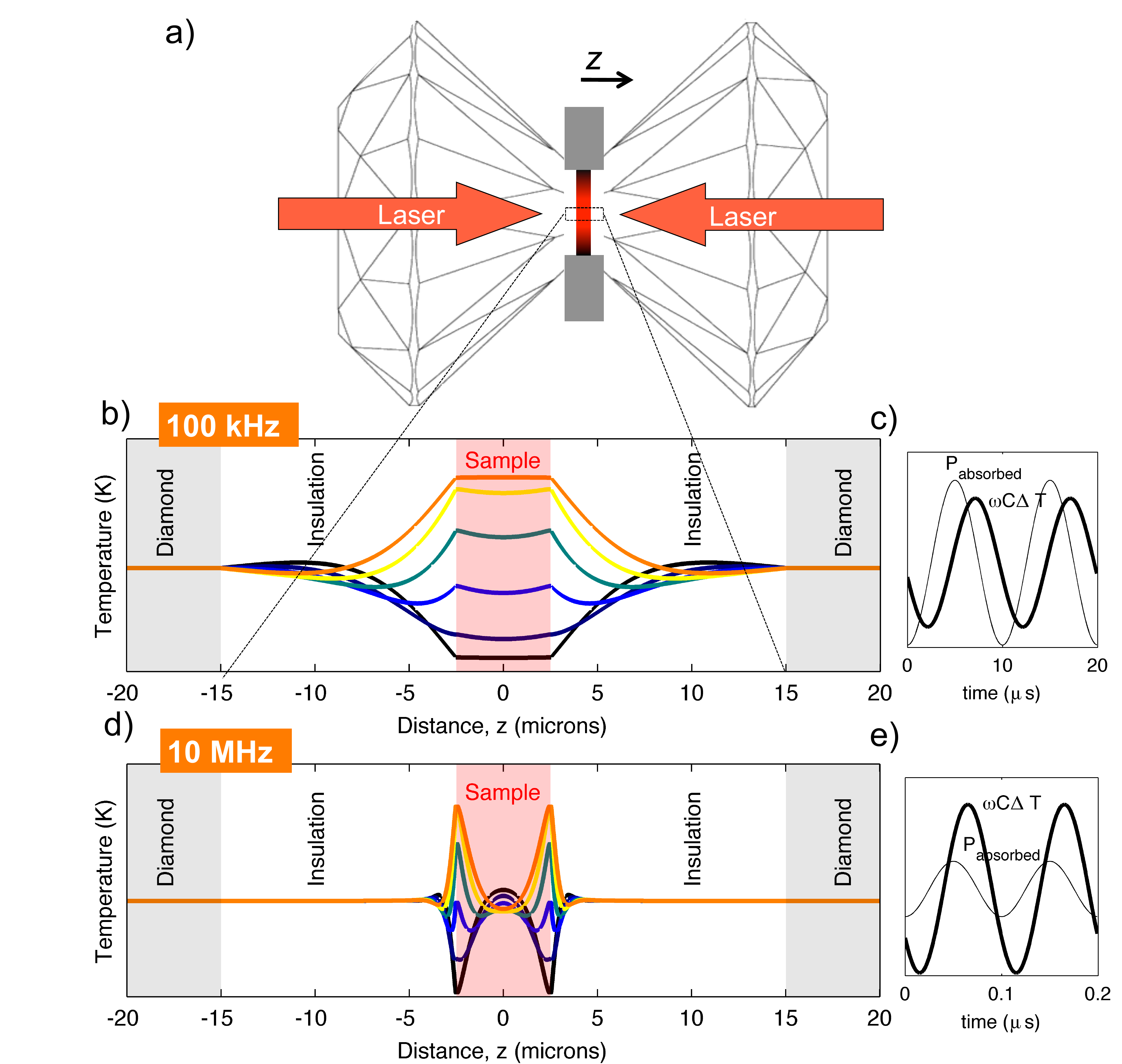}
\caption{(S/S) Schematic of a laser-based experiment for qualitative high pressure-high temperature AC calorimetry. (a) Two heating lasers are incident on the two sides of a sample (not to scale: sample thickness exaggerated in figure). (b,d) Axial temperature profiles, $T(t,z)$, at six times during half of a heating cycle at (b) 100 kHz and (d) 10 MHz frequency of laser modulation. (c,e) Surface temperature oscillations compared to power oscillations during two cycles. The temperature responses to 100 kHz and 10 MHz power modulation appear (c) damped and (e) amplified compared to the power deposited in the sample.}
\label{fig:LH_schematic}
\end{figure*}
\section{}
The main text shows that while laser heating of metals at high frequencies allows for \textit{relative} measurements of heat capacity, no direct measurement of the absolute heat capacity is possible. To give intuition for why no absolute measurement is possible, we present an example of heat deposition and flow during the laser-heating of metals. Figs. \ref{fig:LH_schematic}a-c show that if heat is absorbed at the surface of a 5 $\mu$m-thick metal sample surrounded by 12.5 $\mu$m of KBr insulation on each side, at 100 kHz frequency, the temperature oscillations resulting from laser heating are relatively small due to the large amount of heat that diffuses into the insulation. This causes the total heat capacity that would be measured, $C_\total = \frac{p_\omega}{\omega T_\omega}$, to be large relative to the heat capacity of sample itself, $C$. At 10 MHz, the opposite is true: Figs. \ref{fig:LH_schematic}d-e show that the measured heat capacity would be smaller than the heat capacity of the whole sample since temperature oscillates by more than expected ($\omega C T_\omega > p_\omega$) due to the small penetration depth of the lasers and the limited thermal diffusion into the sample; only the surface of the sample is heated and it has a smaller heat capacity than the full sample. These two examples (100 kHz and 10 MHz) suggest that there is no way to laser-heat metals slowly enough so that they internally equilibrate but fast enough so that little heat is lost to the surroundings.
  
\section{}
\label{apx:3w}

The third harmonic technique provides a way of measuring the amplitude of temperature oscillations based on the third harmonic of voltage created by the temperature oscillations that feed back into voltage oscillations due to the sample's temperature-dependence of resistance.\cite{Kraftmakher2004} Here, we assume that a purely sinusoidal current source ($I = I_\omega \sin(\omega t)$) drives current through the sample. In Part II, we assume a voltage source that is buffered by series resistors rather than the current source assumed here, resulting in a more complicated forms of most equations. Ohm's law gives the voltage:
\begin{eqnarray*}
V &=& IR_{\sam} \\
&=& I_{\omega} \sin(\omega t)R_{\sam} \\
&\approx& V_{\omega}\sin(\omega t) 
\end{eqnarray*}
for some number $V_{\omega}$, which is the first Fourier component of $V$. The approximation is due to the fact that the sample resistance, $R_{\sam}$, is not constant in time. The power deposited in the sample is,
\begin{equation}
p = IV = I_{\omega} ^2 R_{\sam}\sin^2(\omega t) = I_{\omega}^2R_{\sam}(1-\cos(2\omega t))/2
\label{eqn:power}
\end{equation}

The oscillatory term in the expression for power causes a second harmonic temperature oscillation, calculated by equating the heat deposited with change in internal energy. 
\[
\int{-\frac{1}{2}I_{\omega}^2R_{\sam}\cos(2\omega t) dt} = \rho_{\sam}c_{\sam}Vol_{\sam} T_\omega
\]
where $\Delta T$ is the deviation from the steady state background temperature and $Vol$ is the volume of sample. We have assumed adiabatic heating, but note that magnitudes of adiabatic temperature oscillation can be converted to non-adiabatic values by dividing by the values of $C_{meas}/C$ that are shown in many figures. 
Integrating and rearranging,
\begin{eqnarray}
\Delta T &=& -I_\omega ^2 R \sin(2\omega t)/(4\omega\rho_{\sam}c_{\sam}Vol) \nonumber \\
&=& -T_{2\omega} \sin(2\omega t) \nonumber
\label{eqn:DeltaT}
\end{eqnarray}
for the amplitude of temperature oscillation,
\begin{equation}
T_{2\omega} = I_\omega^2 R/(4\omega \rho_{\sam} c_{\sam} Vol)
\label{eqn:T2w}
\end{equation}

This second harmonic temperature variation feeds back into a modulated voltage that contains a third harmonic component, assuming a non-zero temperature-dependence of sample resistance, $d\textrm{log}R/dT$.
\begin{eqnarray*}
V &=& V_{\omega} \sin(\omega t) + \Delta T \cdot I \cdot R_{\sam} \cdot d\textrm{log}R/dT \\
&=& V_{\omega} \sin(\omega t) - T_{2\omega} I_\omega R_{\sam} d\textrm{log}R/dT \cdot \sin(2\omega t) \sin(\omega t) \\
&=& V_\omega \sin(\omega t) +\frac{1}{2}T_{2\omega}I_\omega R_{\sam} d\textrm{log}R/dT (\cos(3\omega t) - \cos(\omega t)) \\
&=& V_\omega \sin(\omega t) + V_{3\omega}\cos(3\omega t) - V_{3\omega}\cos(\omega t)
\end{eqnarray*}
for

\begin{equation}
V_{3\omega} = \frac{1}{2} T_{2\omega} I_\omega R_{\sam} d\textrm{log}R/dT
\label{eqn:V3w_part1}
\end{equation}

Rearranging to solve for $T_{2\omega}$, we conclude that by measuring the amplitude of third harmonic voltage oscillations, the current, resistance and temperature coefficient of resistance (by a preliminary resistance experiment using DC power, for example), the amplitude of temperature oscillations can be inferred:

\begin{equation}
T_{2\omega} = \frac{2 V_{3\omega}}{ I_\omega R_{\sam} d\textrm{log}R/dT} 
\label{eqn:T2w_part1}
\end{equation}
  
\section{} \label{apx:phase-shift}

\begin{figure}
\begin{center}
\includegraphics[width=3.5in]{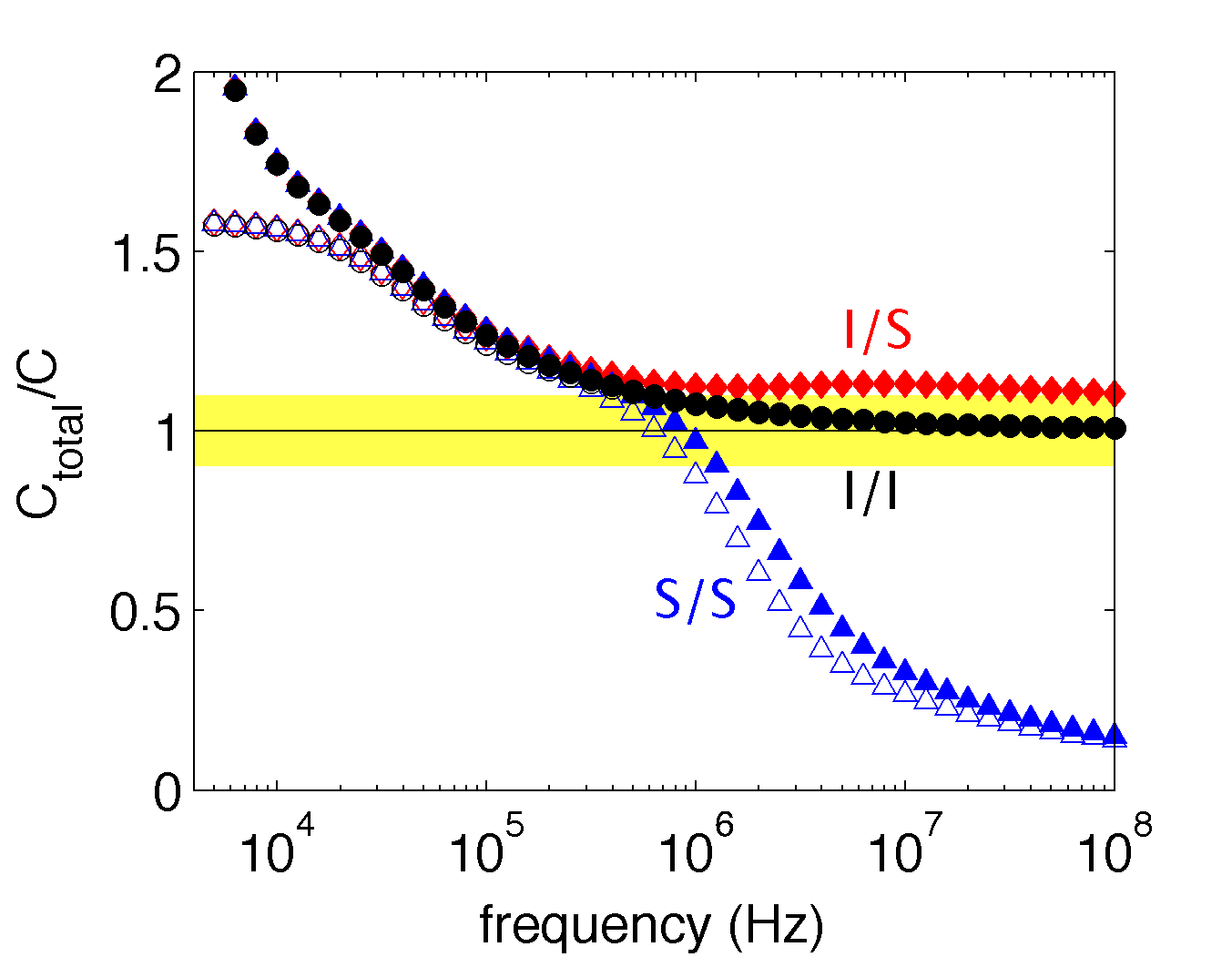} 
\caption{Total heat capacity of sample plus addenda according to two models in which the phase shift due to thermal diffusion into the addenda is taken into account (open symbols, using $\frac{p_\omega}{\omega T_\omega \csc(\phi)}$) and is not taken into account (closed symbols, using $\frac{p_\omega}{\omega T_\omega}$). All points are normalized by sample heat capacity per unit area ($C = \rho_{\sam}c_{\sam}d_{\sam}$) and plotted as a function of the frequency of heating modulation. Black circles indicate that temperature is measured internally in our reference experiment, red diamonds indicate that temperature is measured at the surface ($\delta_{\skin} = 100$ nm), and blue triangles indicate that both temperature measurement and heat deposition take place in the surface ($\delta_{\skin} = 100$ nm). The yellow band marks $< 10 \%$ error in the measured heat capacity.}
\label{fig:int_surf_csc}
\end{center}
\end{figure}

By measuring the phase shift between the heat-source and temperature oscillation, in addition to their amplitudes, it is possible to correct for heat flow into the diamonds anvils (i.e. the constant-temperature boundary imposed in our simulations) using the normal equations of modulation calorimetry. Correction for heat flow into the insulation would require a more complex model because different layers of the insulation respond to sample temperature oscillations with different characteristic times ($\tau \approx \Delta z^2/D_{\ins}$ for the layer that is a distance $\Delta z$ from the sample surface). To account for the diamonds, we combine Eqs. (\ref{eqn:basic}), (\ref{eqn:phase}), and the identity $\csc(\phi) = \sqrt{1+1/\tan^2(\phi)}$, to eliminate the effect of the conductive link through the insulation to the diamond, $K_b$:
\[ T_\omega \csc(\phi) = \frac{p_\omega}{\sqrt{C^2\omega^2 + K_b^2}}\sqrt{1+K_b^2/C^2\omega^2} = \frac{p_\omega}{C\omega}\]
Solving for the total heat capacity of sample plus addenda that would be measured using these equations,
\[ C_\total = \frac{p_\omega}{\omega T_\omega \csc(\phi)} \] 
The total heat capacities that result are shown in Fig. \ref{fig:int_surf_csc}, along with the heat capacities that would be measured assuming $K_b = 0$ (as in the main text). The difference between the two types of total measured heat capacities is the correction for the phase shift ($\phi$) being less than $90^\circ$, and it is significant at frequencies less than 10 kHz, but not at higher frequencies. The reason is simple: the equations assume a disjoint model of sample and thermal bath which describes heat losses to the diamonds but not losses to the insulation that abuts the sample. Hence, the only heat loss that can be corrected for is loss to the diamonds, which becomes significant at $f < D_{\ins}/d_{\ins}^2 = 25$ kHz.

\section{} \label{apx:1sided}
In order to compare the experiments modeled here with typical calorimetry experiments, we alter our simulations in one way: we separate the heater from the temperature probe. For large samples with good insulation, quantitative calorimetry data can be extracted. Moreover, a benefit of separating heater and thermometer is to enable confirmation that the sample itself thermally equilibrates. The challenge is to find a timescale long enough so that the sample internally equilibrates, but short enough so that little heat is lost to the surroundings. Unfortunately, no such timescale exists for diamond-cell sample chambers. The result is that our simulations show no way to make accurate absolute calorimetric measurements when one side of a metal is heated and the temperature is measured on the other side inside a diamond cell.

The same modeling scheme is used here, except the heating source is not given by Eq. (\ref{eqn:Q}), but rather by the following one-sided heater:
\[
Q(z,t) = p_0e^{-(z+d_{sam}/2)/\delta_{skin}}\sin(2\pi f t) 
\]
for $-d_{sam}/2 < z < d_{sam}/2$. Temperature is measured at $z = d_{sam}/2$.

\begin{figure}
\begin{center}
\includegraphics[width=3in]{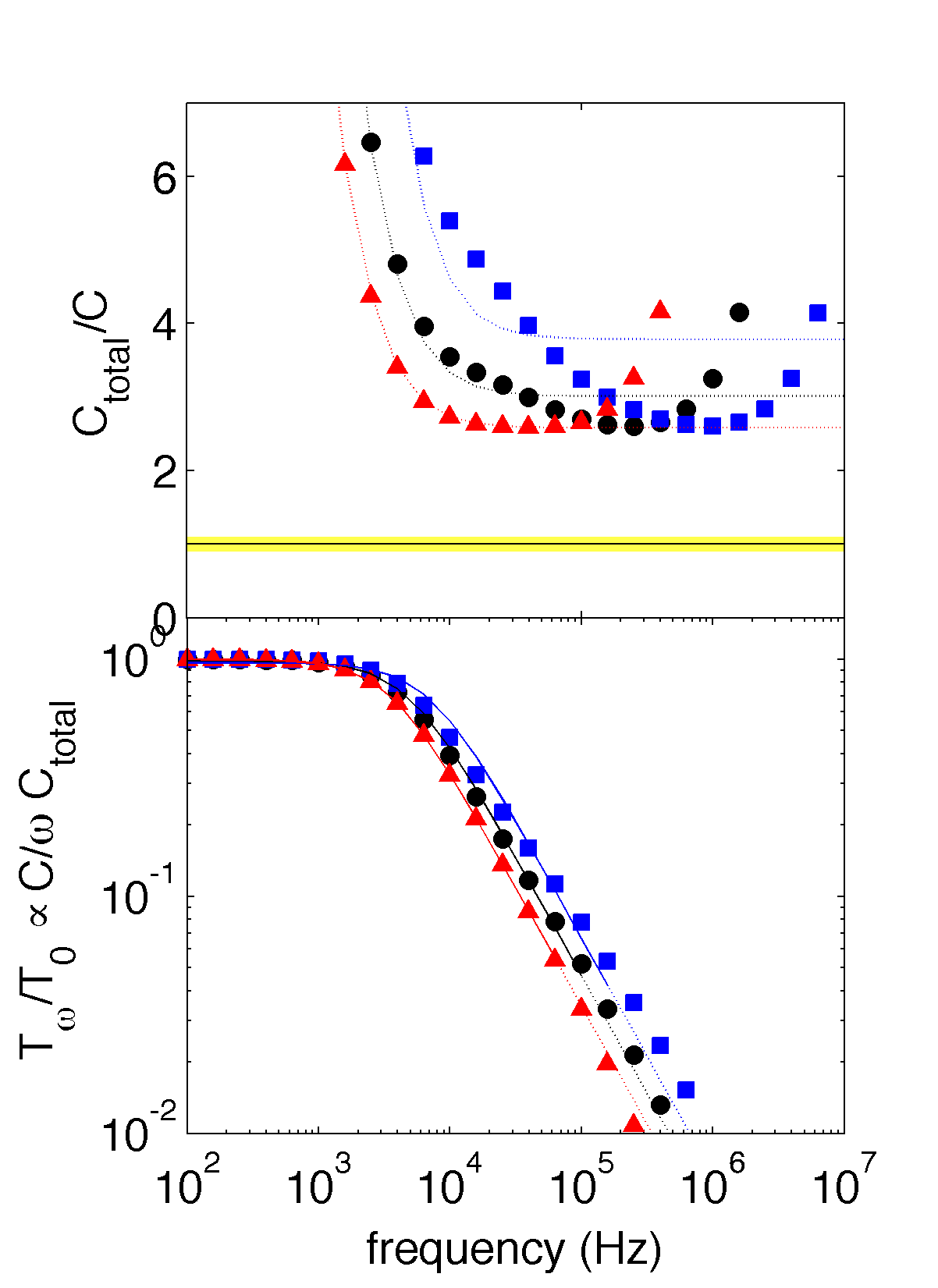} 
\caption{Top: Heat capacity inferred when heating a sample from one side and measuring temperature at the other side. The sample is 5 $\mu$m-thick (black circles), 10 $\mu$m-thick (red triangles) or 2.5 $\mu$m-thick (blue squares), while the full distance between the diamond anvils is 30 $\mu$m. Dotted curves are fits from the bottom panel. Bottom: Normalized values of temperature oscillation amplitude for a fixed amplitude of power oscillation, assuming the same samples as in the top panel. Solid curves are fits of the modeled temperature oscillations between $10^4$ and $10^7$ Hz using Eq. (\ref{eqn:box-model}) and minimizing the variance in log space, $\sum_i (\textrm{log}(\Delta T_i^{\textrm{calc}})-\textrm{log}(\Delta T_i))^2$, while dotted lines are the same fits extended to higher frequencies.}
\label{fig:Bouquet_fits}
\end{center}
\end{figure}

Measured specific heats are plotted in Fig. \ref{fig:Bouquet_fits} for three thicknesses of samples inside a sample chamber whose total thickness is 30 $\mu$m from diamond to diamond. In all cases, the minimum measured heat capacity is over 150\% larger than the true heat capacity due to the large addenda at low frequencies and the lack of thermal equilibration between heater and temperature probe at high frequencies. 

Despite the 150\% error in measurement of the sample's heat capacity, measured temperature variations can appear to fit the equation for non-adiabatic calorimetry that assumes disjoint components (sample, thermal bath, and negligibly small heater and thermometer) connected by a thermal link, $K_b$:  \footnote{This is Eq. (2.3c) of Ref. \onlinecite{Kraftmakher2004}, Eq. (1) of Ref. \onlinecite{Bouquet2000} and Eq. (11) of Ref. \onlinecite{Sullivan1968} with $\tau_1 = C/K_b$, $\tau_2 \to 0$, and $K_b/K_s \to 0$}
\begin{equation}
T_\omega = \frac{p}{\sqrt{C^2\omega^2 + K_b^2}}
\label{eqn:box-model}
\end{equation}
where $\omega = 2\pi f$, $C$ is the heat capacity of the sample and addenda, and $K_b$ is the thermal conductance of the link to the temperature bath ($K$ in Fig. 1a of Ref. \onlinecite{Bouquet2000}; $Q'$ in eqn 2.3c of Ref. \onlinecite{Kraftmakher2004}). Fig. \ref{fig:Bouquet_fits} shows simulation results for three thicknesses of sample, along with the simple model, where $\textrm{log}(T_\omega)$ has been fitted over the range $10^2$ to $\sim 10^5$ Hz. Good fits result from simulated data when $d_\sam$ is 5 or 10 $\mu$m, but even the 2.5 $\mu$m thick sample can be fitted approximately.

\section{}  \label{apx:crank}

In all heat-flow calculations, we use the Crank-Nicholson method to solve the one-dimensional heat equation with Dirichlet boundary conditions at the diamond surfaces and a Neumann (no flow) boundary condition at the center of the sample when heating is symmetric. The equation that we use to solve for temperatures at time step $n+1$ based on temperatures at time step $n$ is 
\begin{equation}
\textbf{A}T^{n+1} = \textbf{B}T^n + D
\label{eqn:crank}
\end{equation}
where $T^{n+1}$ and $T^n$ are column vectors describing the temperature at all pixels, $\textbf{A}$ and $\textbf{B}$ are matrices that account for thermal diffusion, and $D$ is a column vector that accounts for the heating source. By writing out explicit and implicit formulations of the discretized heat equation with a source term, taking special care at the surface of the sample (we assume the surface pixel has the density and heat capacity of the sample but is connected to the next pixel by a region of conductivity $k_\ins$), and rearranging to match the form of Eq. (\ref{eqn:crank}), we find that $\textbf{A}$ is a tri-diagonal matrix whose entries above, on, and below the diagonal are

\begin{widetext}
\[
\begin{array}{p{3 cm} cccccccccc}
\textrm{above diagonal}: &-2r_{s} & -r_s&...&  -r_{s} & -r_{s}\frac{k_{\ins}}{k_{\sam}} & -r_{i}&...&  -r_{i} & -r_{i} & \\
\textrm{on diagonal}: &2r_{s}+1 & 2r_s + 1&... & 2r_{s} + 1 & 1+r_{s}(1+\frac{k_{\ins}}{k_{\sam}}) & 2r_{i} + 1 & ... & 2r_{i} + 1 & 2r_{i} + 1 &1 \\
\textrm{below diagonal}: & & -r_{s} & ... &-r_s& -r_s & -r_i & ...&-r_i  &-r_i & 0 \\
\end{array} 
 \nonumber 
\]
\end{widetext}

for 
\[
r_{s} = \frac{k_{\sam}\Delta t}{2 c_{\sam}\rho_{\sam} \Delta z^2}
\]
and 
\[
r_{i} = \frac{k_{\ins}\Delta t}{2 c_{\ins}\rho_{\ins} \Delta z^2}
\]
The other large matrices are given by
\[\textbf{B} = 2-\textbf{A} \]
where 2 is the 2 times the identity matrix and by
\begin{widetext}
\[ D = \frac{\Delta t}{2}p_0\cdot \exp\left((d_{\sam}/2-\left[\begin{array}{c}
0\\
\Delta z\\
2\Delta z\\
...\\
d_{\sam}/2\\
0\\
...\\
0 \\
\end{array} \right])/\delta_{\skin}\right)\Big(\sin(2\pi f t) + \sin(2\pi f (t+\Delta t)\Big)
\]
\end{widetext}
We use a mesh, $\Delta z$, of 1 nm and a timestep, $\Delta t$, of $0.01/f$.

For one-sided heating (as in Appendix C), we increase the size of matrices $\textbf{A}$ and $\textbf{B}$ by deleting the top row of each and concatenating each with a flipped image of itself. We perform a similar operation to the vector $D$, and modify it to describe one-sided heating that peaks at $z = -d_{\sam}/2$ and decreases exponentially through the width of the sample.

\section{} \label{apx:2thicknesses}
\begin{figure}
\begin{center}
\includegraphics[width=3.5in]{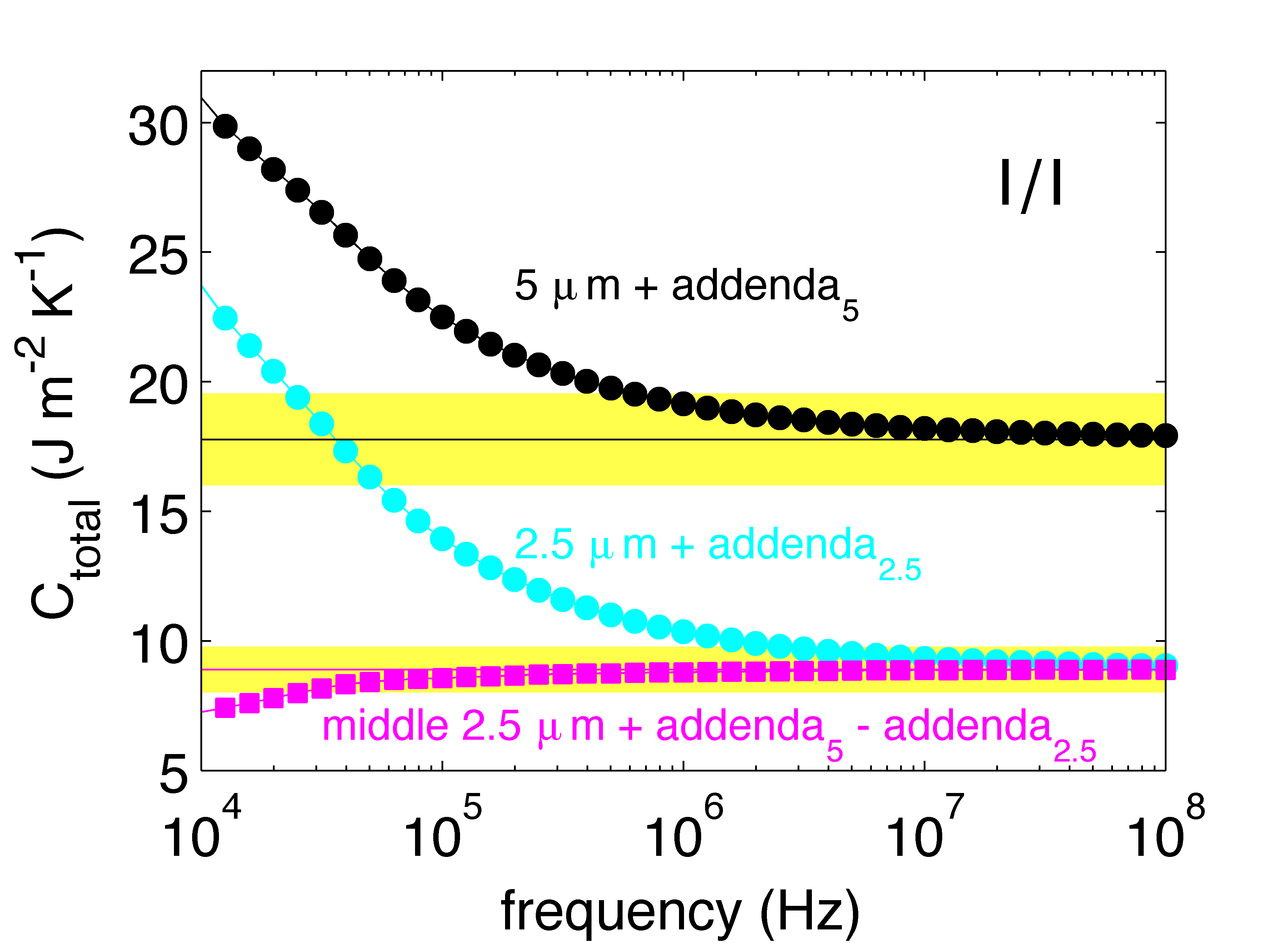}
\caption{Heat capacities that can be inferred by measuring two resistively heated samples with thicknesses of 5 $\mu$m and 2.5 $\mu$m, assuming internal temperature measurement and the reference geometry and material properties of Table \ref{table:mat_props}. The total heat capacity of sample plus addenda per unit area, $C_\total = p_{2\omega}/(2\omega  T_{2\omega})$, is normalized by the sample heat capacity per unit area, $\rho_{\sam}c_{\sam}d_{\sam}$, for the 2.5 $\mu$m-thick sample (cyan circles) and the 1 $\mu$m-thick sample (black circles). By combining both measurements, we infer the heat capacity of the middle 2.5 $\mu$m in a way that nearly eliminates contribution from the addenda: pink squares show $(C_{\total,5}- C_{\total,2.5})$ normalized by the heat capacity of a 2.5 $\mu$m-thick sample. The yellow bands marks $< 10 \%$ error in heat capacity measurement.}
\label{fig:2thicknesses}
\end{center}
\end{figure}

One way to improve accuracy or verify accuracy of heat capacity measurements is to repeat an experiment with two or more thicknesses of sample. Here we compare the heat capacity that would be inferred if the sample were 5 or 2.5 $\mu$m thick, but all other properties identical to the reference simulation of the main text. By subtracting the measured heat capacity of the thinner sample from that of the thicker sample, Fig. \ref{fig:2thicknesses} shows that addenda approximately cancel, leaving the heat capacity of a 2.5 $\mu$m thick sample plus a small addendum. The frequency required to achieve 10\% accuracy using this ``two-thicknesses'' trick is 20 kHz, 1.5 orders of magnitude slower than the 600 kHz frequency required when a single sample of 5 $\mu$m-thickness is measured.

\begin{acknowledgements}
We thank Sam Weir, Jon Eggert, Jaijeet Roychowdhury, Gernot Pottlacher and Claus Cagran for stimulating discussion. Support for Z.M.G. was provided by the NSF GRFP and by CDAC.
\end{acknowledgements}

\nocite{*}
\bibliography{AC_Cal1_v2}

\begin{thebibliography}{36}%
\makeatletter
\providecommand \@ifxundefined [1]{%
 \@ifx{#1\undefined}
}%
\providecommand \@ifnum [1]{%
 \ifnum #1\expandafter \@firstoftwo
 \else \expandafter \@secondoftwo
 \fi
}%
\providecommand \@ifx [1]{%
 \ifx #1\expandafter \@firstoftwo
 \else \expandafter \@secondoftwo
 \fi
}%
\providecommand \natexlab [1]{#1}%
\providecommand \enquote  [1]{``#1''}%
\providecommand \bibnamefont  [1]{#1}%
\providecommand \bibfnamefont [1]{#1}%
\providecommand \citenamefont [1]{#1}%
\providecommand \href@noop [0]{\@secondoftwo}%
\providecommand \href [0]{\begingroup \@sanitize@url \@href}%
\providecommand \@href[1]{\@@startlink{#1}\@@href}%
\providecommand \@@href[1]{\endgroup#1\@@endlink}%
\providecommand \@sanitize@url [0]{\catcode `\\12\catcode `\$12\catcode
  `\&12\catcode `\#12\catcode `\^12\catcode `\_12\catcode `\%12\relax}%
\providecommand \@@startlink[1]{}%
\providecommand \@@endlink[0]{}%
\providecommand \url  [0]{\begingroup\@sanitize@url \@url }%
\providecommand \@url [1]{\endgroup\@href {#1}{\urlprefix }}%
\providecommand \urlprefix  [0]{URL }%
\providecommand \Eprint [0]{\href }%
\providecommand \doibase [0]{http://dx.doi.org/}%
\providecommand \selectlanguage [0]{\@gobble}%
\providecommand \bibinfo  [0]{\@secondoftwo}%
\providecommand \bibfield  [0]{\@secondoftwo}%
\providecommand \translation [1]{[#1]}%
\providecommand \BibitemOpen [0]{}%
\providecommand \bibitemStop [0]{}%
\providecommand \bibitemNoStop [0]{.\EOS\space}%
\providecommand \EOS [0]{\spacefactor3000\relax}%
\providecommand \BibitemShut  [1]{\csname bibitem#1\endcsname}%
\let\auto@bib@innerbib\@empty
\bibitem [{\citenamefont {Pais}(1979)}]{Pais1979}%
  \BibitemOpen
  \bibfield  {author} {\bibinfo {author} {\bibfnamefont {A.}~\bibnamefont
  {Pais}},\ }\href@noop {} {\bibfield  {journal} {\bibinfo  {journal} {Rev.
  Mod. Phys.}\ }\textbf {\bibinfo {volume} {51}},\ \bibinfo {pages} {863}
  (\bibinfo {year} {1979})}\BibitemShut {NoStop}%
\bibitem [{\citenamefont {Debye}(1912)}]{Debye1912}%
  \BibitemOpen
  \bibfield  {author} {\bibinfo {author} {\bibfnamefont {P.}~\bibnamefont
  {Debye}},\ }\href@noop {} {\bibfield  {journal} {\bibinfo  {journal} {Annalen
  der Physik}\ }\textbf {\bibinfo {volume} {344}},\ \bibinfo {pages} {789}
  (\bibinfo {year} {1912})}\BibitemShut {NoStop}%
\bibitem [{\citenamefont {Donnelly}(1995)}]{Donnelly1995}%
  \BibitemOpen
  \bibfield  {author} {\bibinfo {author} {\bibfnamefont {R.~J.}\ \bibnamefont
  {Donnelly}},\ }\href@noop {} {\bibfield  {journal} {\bibinfo  {journal}
  {Physics Today}\ }\textbf {\bibinfo {volume} {48}},\ \bibinfo {pages} {30}
  (\bibinfo {year} {1995})}\BibitemShut {NoStop}%
\bibitem [{\citenamefont {Lipa}\ \emph {et~al.}(1996)\citenamefont {Lipa},
  \citenamefont {Swanson}, \citenamefont {Nissen}, \citenamefont {Chui},\ and\
  \citenamefont {Israelsson}}]{Lipa1996}%
  \BibitemOpen
  \bibfield  {author} {\bibinfo {author} {\bibfnamefont {J.~A.}\ \bibnamefont
  {Lipa}}, \bibinfo {author} {\bibfnamefont {D.~R.}\ \bibnamefont {Swanson}},
  \bibinfo {author} {\bibfnamefont {J.~A.}\ \bibnamefont {Nissen}}, \bibinfo
  {author} {\bibfnamefont {T.~C.}\ \bibnamefont {Chui}}, \ and\ \bibinfo
  {author} {\bibfnamefont {U.~E.}\ \bibnamefont {Israelsson}},\ }\href
  {http://www.ncbi.nlm.nih.gov/pubmed/10061591} {\bibfield  {journal} {\bibinfo
   {journal} {Phys. Rev. Lett.}\ }\textbf {\bibinfo {volume} {76}},\ \bibinfo
  {pages} {944} (\bibinfo {year} {1996})}\BibitemShut {NoStop}%
\bibitem [{\citenamefont {Lange}\ and\ \citenamefont
  {Carmichael}(1990)}]{Lange1990}%
  \BibitemOpen
  \bibfield  {author} {\bibinfo {author} {\bibfnamefont {R.~L.}\ \bibnamefont
  {Lange}}\ and\ \bibinfo {author} {\bibfnamefont {I.~S.~E.}\ \bibnamefont
  {Carmichael}},\ }\href@noop {} {\bibfield  {journal} {\bibinfo  {journal}
  {Rev. Mineral.}\ }\textbf {\bibinfo {volume} {24}},\ \bibinfo {pages} {25}
  (\bibinfo {year} {1990})}\BibitemShut {NoStop}%
\bibitem [{\citenamefont {Demuer}\ \emph {et~al.}(2000)\citenamefont {Demuer},
  \citenamefont {Marcenat}, \citenamefont {Thomasson}, \citenamefont
  {Calemczuk}, \citenamefont {Salce}, \citenamefont {Lejay}, \citenamefont
  {Braithwaite},\ and\ \citenamefont {Flouquet}}]{Demuer2000}%
  \BibitemOpen
  \bibfield  {author} {\bibinfo {author} {\bibfnamefont {A.}~\bibnamefont
  {Demuer}}, \bibinfo {author} {\bibfnamefont {C.}~\bibnamefont {Marcenat}},
  \bibinfo {author} {\bibfnamefont {J.}~\bibnamefont {Thomasson}}, \bibinfo
  {author} {\bibfnamefont {R.}~\bibnamefont {Calemczuk}}, \bibinfo {author}
  {\bibfnamefont {B.}~\bibnamefont {Salce}}, \bibinfo {author} {\bibfnamefont
  {P.}~\bibnamefont {Lejay}}, \bibinfo {author} {\bibfnamefont
  {D.}~\bibnamefont {Braithwaite}}, \ and\ \bibinfo {author} {\bibfnamefont
  {J.}~\bibnamefont {Flouquet}},\ }\href@noop {} {\bibfield  {journal}
  {\bibinfo  {journal} {J. Low Temp. Phys.}\ }\textbf {\bibinfo {volume}
  {120}},\ \bibinfo {pages} {245} (\bibinfo {year} {2000})}\BibitemShut
  {NoStop}%
\bibitem [{\citenamefont {Fernandez-Pa{\~{n}}ella}\ \emph
  {et~al.}(2011)\citenamefont {Fernandez-Pa{\~{n}}ella}, \citenamefont
  {Braithwaite}, \citenamefont {Salce}, \citenamefont {Lapertot},\ and\
  \citenamefont {Flouquet}}]{Fernandez-Panella2011}%
  \BibitemOpen
  \bibfield  {author} {\bibinfo {author} {\bibfnamefont {A.}~\bibnamefont
  {Fernandez-Pa{\~{n}}ella}}, \bibinfo {author} {\bibfnamefont
  {D.}~\bibnamefont {Braithwaite}}, \bibinfo {author} {\bibfnamefont
  {B.}~\bibnamefont {Salce}}, \bibinfo {author} {\bibfnamefont
  {G.}~\bibnamefont {Lapertot}}, \ and\ \bibinfo {author} {\bibfnamefont
  {J.}~\bibnamefont {Flouquet}},\ }\href {\doibase 10.1103/PhysRevB.84.134416}
  {\bibfield  {journal} {\bibinfo  {journal} {Phys. Rev. B}\ }\textbf {\bibinfo
  {volume} {84}},\ \bibinfo {pages} {134416} (\bibinfo {year}
  {2011})}\BibitemShut {NoStop}%
\bibitem [{\citenamefont {Wilhelm}, \citenamefont {Revaz},\ and\ \citenamefont
  {Jaccard}(1999)}]{Wilhelm1999}%
  \BibitemOpen
  \bibfield  {author} {\bibinfo {author} {\bibfnamefont {H.}~\bibnamefont
  {Wilhelm}}, \bibinfo {author} {\bibfnamefont {B.}~\bibnamefont {Revaz}}, \
  and\ \bibinfo {author} {\bibfnamefont {D.}~\bibnamefont {Jaccard}},\
  }\href@noop {} {\bibfield  {journal} {\bibinfo  {journal} {Phys. Rev. B}\
  }\textbf {\bibinfo {volume} {59}},\ \bibinfo {pages} {3651} (\bibinfo {year}
  {1999})}\BibitemShut {NoStop}%
\bibitem [{\citenamefont {Bouquet}\ \emph {et~al.}(2000)\citenamefont
  {Bouquet}, \citenamefont {Wang}, \citenamefont {Wilhelm}, \citenamefont
  {Jaccard},\ and\ \citenamefont {Junod}}]{Bouquet2000}%
  \BibitemOpen
  \bibfield  {author} {\bibinfo {author} {\bibfnamefont {F.}~\bibnamefont
  {Bouquet}}, \bibinfo {author} {\bibfnamefont {Y.}~\bibnamefont {Wang}},
  \bibinfo {author} {\bibfnamefont {H.}~\bibnamefont {Wilhelm}}, \bibinfo
  {author} {\bibfnamefont {D.}~\bibnamefont {Jaccard}}, \ and\ \bibinfo
  {author} {\bibfnamefont {A.}~\bibnamefont {Junod}},\ }\href@noop {}
  {\bibfield  {journal} {\bibinfo  {journal} {Solid State Comm.}\ }\textbf
  {\bibinfo {volume} {113}},\ \bibinfo {pages} {367} (\bibinfo {year}
  {2000})}\BibitemShut {NoStop}%
\bibitem [{\citenamefont {Sidorov}, \citenamefont {Thompson},\ and\
  \citenamefont {Fisk}(2010)}]{Sidorov2010}%
  \BibitemOpen
  \bibfield  {author} {\bibinfo {author} {\bibfnamefont {V.~A.}\ \bibnamefont
  {Sidorov}}, \bibinfo {author} {\bibfnamefont {J.~D.}\ \bibnamefont
  {Thompson}}, \ and\ \bibinfo {author} {\bibfnamefont {Z.}~\bibnamefont
  {Fisk}},\ }\href {\doibase 10.1088/0953-8984/22/40/406002} {\bibfield
  {journal} {\bibinfo  {journal} {J. Phys. Cond. Matt.}\ }\textbf {\bibinfo
  {volume} {22}},\ \bibinfo {pages} {406002} (\bibinfo {year}
  {2010})}\BibitemShut {NoStop}%
\bibitem [{\citenamefont {Hicks}\ \emph {et~al.}(2006)\citenamefont {Hicks},
  \citenamefont {Boehly}, \citenamefont {Eggert}, \citenamefont {Miller},
  \citenamefont {Celliers},\ and\ \citenamefont {Collins}}]{hicks2006}%
  \BibitemOpen
  \bibfield  {author} {\bibinfo {author} {\bibfnamefont {D.~G.}\ \bibnamefont
  {Hicks}}, \bibinfo {author} {\bibfnamefont {T.~R.}\ \bibnamefont {Boehly}},
  \bibinfo {author} {\bibfnamefont {J.~H.}\ \bibnamefont {Eggert}}, \bibinfo
  {author} {\bibfnamefont {J.~E.}\ \bibnamefont {Miller}}, \bibinfo {author}
  {\bibfnamefont {P.~M.}\ \bibnamefont {Celliers}}, \ and\ \bibinfo {author}
  {\bibfnamefont {G.~W.}\ \bibnamefont {Collins}},\ }\href {\doibase
  10.1103/PhysRevLett.97.025502} {\bibfield  {journal} {\bibinfo  {journal}
  {Phys. Rev. Lett.}\ }\textbf {\bibinfo {volume} {97}},\ \bibinfo {pages}
  {025502} (\bibinfo {year} {2006})}\BibitemShut {NoStop}%
\bibitem [{\citenamefont {Eggert}\ \emph {et~al.}(2009)\citenamefont {Eggert},
  \citenamefont {Hicks}, \citenamefont {Celliers}, \citenamefont {Bradley},
  \citenamefont {McWilliams}, \citenamefont {Jeanloz}, \citenamefont {Miller},
  \citenamefont {Boehly},\ and\ \citenamefont {Collins}}]{Eggert2009}%
  \BibitemOpen
  \bibfield  {author} {\bibinfo {author} {\bibfnamefont {J.~H.}\ \bibnamefont
  {Eggert}}, \bibinfo {author} {\bibfnamefont {D.~G.}\ \bibnamefont {Hicks}},
  \bibinfo {author} {\bibfnamefont {P.~M.}\ \bibnamefont {Celliers}}, \bibinfo
  {author} {\bibfnamefont {D.~K.}\ \bibnamefont {Bradley}}, \bibinfo {author}
  {\bibfnamefont {R.~S.}\ \bibnamefont {McWilliams}}, \bibinfo {author}
  {\bibfnamefont {R.}~\bibnamefont {Jeanloz}}, \bibinfo {author} {\bibfnamefont
  {J.~E.}\ \bibnamefont {Miller}}, \bibinfo {author} {\bibfnamefont {T.~R.}\
  \bibnamefont {Boehly}}, \ and\ \bibinfo {author} {\bibfnamefont {G.~W.}\
  \bibnamefont {Collins}},\ }\href {\doibase 10.1038/nphys1438} {\bibfield
  {journal} {\bibinfo  {journal} {Nature Phys.}\ }\textbf {\bibinfo {volume}
  {6}},\ \bibinfo {pages} {40} (\bibinfo {year} {2009})}\BibitemShut {NoStop}%
\bibitem [{\citenamefont {Hazen}\ and\ \citenamefont
  {Navrotsky}(1996)}]{Hazen1996}%
  \BibitemOpen
  \bibfield  {author} {\bibinfo {author} {\bibfnamefont {R.~M.}\ \bibnamefont
  {Hazen}}\ and\ \bibinfo {author} {\bibfnamefont {A.}~\bibnamefont
  {Navrotsky}},\ }\href@noop {} {\bibfield  {journal} {\bibinfo  {journal} {Am.
  Mineral.}\ }\textbf {\bibinfo {volume} {81}},\ \bibinfo {pages} {1021}
  (\bibinfo {year} {1996})}\BibitemShut {NoStop}%
\bibitem [{\citenamefont {Petrova}\ and\ \citenamefont
  {Stishov}(2012)}]{Petrova2012}%
  \BibitemOpen
  \bibfield  {author} {\bibinfo {author} {\bibfnamefont {A.~E.}\ \bibnamefont
  {Petrova}}\ and\ \bibinfo {author} {\bibfnamefont {S.~M.}\ \bibnamefont
  {Stishov}},\ }\href {\doibase 10.1103/PhysRevB.86.174407} {\bibfield
  {journal} {\bibinfo  {journal} {Phys. Rev. B}\ }\textbf {\bibinfo {volume}
  {86}},\ \bibinfo {pages} {174407} (\bibinfo {year} {2012})}\BibitemShut
  {NoStop}%
\bibitem [{\citenamefont {Sullivan}\ and\ \citenamefont
  {Seidel}(1968)}]{Sullivan1968}%
  \BibitemOpen
  \bibfield  {author} {\bibinfo {author} {\bibfnamefont {P.~F.}\ \bibnamefont
  {Sullivan}}\ and\ \bibinfo {author} {\bibfnamefont {G.}~\bibnamefont
  {Seidel}},\ }\href@noop {} {\bibfield  {journal} {\bibinfo  {journal} {Phys.
  Rev.}\ }\textbf {\bibinfo {volume} {173}},\ \bibinfo {pages} {679} (\bibinfo
  {year} {1968})}\BibitemShut {NoStop}%
\bibitem [{\citenamefont {Kraftmakher}(2004)}]{Kraftmakher2004}%
  \BibitemOpen
  \bibfield  {author} {\bibinfo {author} {\bibfnamefont {Y.}~\bibnamefont
  {Kraftmakher}},\ }\href@noop {} {\emph {\bibinfo {title} {{Modulation
  Calorimetry: Theory and Applications}}}}\ (\bibinfo  {publisher} {Springer},\
  \bibinfo {address} {Berlin},\ \bibinfo {year} {2004})\ p.\ \bibinfo {pages}
  {284}\BibitemShut {NoStop}%
\bibitem [{Note1()}]{Note1}%
  \BibitemOpen
  \bibinfo {note} {This is Eq. (2.3c) of Ref. \protect \rev@citealpnum
  {Kraftmakher2004} with $Q'$ instead of $K_b$, Eq. (2) of Ref. \protect
  \rev@citealpnum {Baloga1977} with $\Gamma $ instead of $K_b$, Eq. (1) of Ref.
  \protect \rev@citealpnum {Bouquet2000}, or Eq. (11) of Ref. \protect
  \rev@citealpnum {Sullivan1968} with $\tau _1 = C/K_b$, $\tau _2 \to 0$, and
  $K_b/K_s \to 0$.}\BibitemShut {Stop}%
\bibitem [{\citenamefont {Baloga}\ and\ \citenamefont
  {Garland}(1977)}]{Baloga1977}%
  \BibitemOpen
  \bibfield  {author} {\bibinfo {author} {\bibfnamefont {J.~D.}\ \bibnamefont
  {Baloga}}\ and\ \bibinfo {author} {\bibfnamefont {C.~W.}\ \bibnamefont
  {Garland}},\ }\href@noop {} {\bibfield  {journal} {\bibinfo  {journal} {Rev.
  Sci. Instrum.}\ }\textbf {\bibinfo {volume} {48}},\ \bibinfo {pages} {105}
  (\bibinfo {year} {1977})}\BibitemShut {NoStop}%
\bibitem [{Note2()}]{Note2}%
  \BibitemOpen
  \bibinfo {note} {Ref. \protect \rev@citealpnum {Baloga1977} uses an analytic
  expression for an addenda contribution to heat capacity at variable frequency
  and shows that it fits their data.}\BibitemShut {Stop}%
\bibitem [{\citenamefont {Geballe}\ and\ \citenamefont
  {Jeanloz}(2012)}]{Geballe2012}%
  \BibitemOpen
  \bibfield  {author} {\bibinfo {author} {\bibfnamefont {Z.~M.}\ \bibnamefont
  {Geballe}}\ and\ \bibinfo {author} {\bibfnamefont {R.}~\bibnamefont
  {Jeanloz}},\ }\href {\doibase 10.1063/1.4729905} {\bibfield  {journal}
  {\bibinfo  {journal} {J. Appl. Phys.}\ }\textbf {\bibinfo {volume} {111}},\
  \bibinfo {pages} {123518} (\bibinfo {year} {2012})}\BibitemShut {NoStop}%
\bibitem [{\citenamefont {Weir}\ \emph {et~al.}(2009)\citenamefont {Weir},
  \citenamefont {Jackson}, \citenamefont {Falabella}, \citenamefont
  {Samudrala},\ and\ \citenamefont {Vohra}}]{Weir2009a}%
  \BibitemOpen
  \bibfield  {author} {\bibinfo {author} {\bibfnamefont {S.~T.}\ \bibnamefont
  {Weir}}, \bibinfo {author} {\bibfnamefont {D.~D.}\ \bibnamefont {Jackson}},
  \bibinfo {author} {\bibfnamefont {S.}~\bibnamefont {Falabella}}, \bibinfo
  {author} {\bibfnamefont {G.}~\bibnamefont {Samudrala}}, \ and\ \bibinfo
  {author} {\bibfnamefont {Y.~K.}\ \bibnamefont {Vohra}},\ }\href {\doibase
  10.1063/1.3069286} {\bibfield  {journal} {\bibinfo  {journal} {Rev. Sci.
  Instrum.}\ }\textbf {\bibinfo {volume} {80}},\ \bibinfo {pages} {013905}
  (\bibinfo {year} {2009})}\BibitemShut {NoStop}%
\bibitem [{\citenamefont {Zha}, \citenamefont {Bassett},\ and\ \citenamefont
  {Shim}(2004)}]{Zha2004}%
  \BibitemOpen
  \bibfield  {author} {\bibinfo {author} {\bibfnamefont {C.-S.}\ \bibnamefont
  {Zha}}, \bibinfo {author} {\bibfnamefont {W.~a.}\ \bibnamefont {Bassett}}, \
  and\ \bibinfo {author} {\bibfnamefont {S.-H.}\ \bibnamefont {Shim}},\ }\href
  {\doibase 10.1063/1.1765752} {\bibfield  {journal} {\bibinfo  {journal} {Rev.
  Sci. Instrum.}\ }\textbf {\bibinfo {volume} {75}},\ \bibinfo {pages} {2409}
  (\bibinfo {year} {2004})}\BibitemShut {NoStop}%
\bibitem [{\citenamefont {Komabayashi}\ \emph {et~al.}(2009)\citenamefont
  {Komabayashi}, \citenamefont {Fei}, \citenamefont {Meng},\ and\ \citenamefont
  {Prakapenka}}]{Komabayashi2009}%
  \BibitemOpen
  \bibfield  {author} {\bibinfo {author} {\bibfnamefont {T.}~\bibnamefont
  {Komabayashi}}, \bibinfo {author} {\bibfnamefont {Y.}~\bibnamefont {Fei}},
  \bibinfo {author} {\bibfnamefont {Y.}~\bibnamefont {Meng}}, \ and\ \bibinfo
  {author} {\bibfnamefont {V.}~\bibnamefont {Prakapenka}},\ }\href {\doibase
  10.1016/j.epsl.2009.03.025} {\bibfield  {journal} {\bibinfo  {journal} {Earth
  and Planetary Science Letters}\ }\textbf {\bibinfo {volume} {282}},\ \bibinfo
  {pages} {252} (\bibinfo {year} {2009})}\BibitemShut {NoStop}%
\bibitem [{\citenamefont {Melentiev}, \citenamefont {Subbotin},\ and\
  \citenamefont {Balykin}(2001)}]{Melentiev2001}%
  \BibitemOpen
  \bibfield  {author} {\bibinfo {author} {\bibfnamefont {P.~N.}\ \bibnamefont
  {Melentiev}}, \bibinfo {author} {\bibfnamefont {M.~V.}\ \bibnamefont
  {Subbotin}}, \ and\ \bibinfo {author} {\bibfnamefont {V.~I.}\ \bibnamefont
  {Balykin}},\ }\href@noop {} {\bibfield  {journal} {\bibinfo  {journal} {Laser
  Phys.}\ }\textbf {\bibinfo {volume} {11}},\ \bibinfo {pages} {891} (\bibinfo
  {year} {2001})}\BibitemShut {NoStop}%
\bibitem [{\citenamefont {Cahill}(2004)}]{Cahill2004}%
  \BibitemOpen
  \bibfield  {author} {\bibinfo {author} {\bibfnamefont {D.~G.}\ \bibnamefont
  {Cahill}},\ }\href {\doibase 10.1063/1.1819431} {\bibfield  {journal}
  {\bibinfo  {journal} {Rev. Sci. Instrum.}\ }\textbf {\bibinfo {volume}
  {75}},\ \bibinfo {pages} {5119} (\bibinfo {year} {2004})}\BibitemShut
  {NoStop}%
\bibitem [{\citenamefont {Cahill}(1990)}]{Cahill1990}%
  \BibitemOpen
  \bibfield  {author} {\bibinfo {author} {\bibfnamefont {D.~G.}\ \bibnamefont
  {Cahill}},\ }\href@noop {} {\bibfield  {journal} {\bibinfo  {journal} {Rev.
  Sci. Instrum.}\ }\textbf {\bibinfo {volume} {61}},\ \bibinfo {pages} {802}
  (\bibinfo {year} {1990})}\BibitemShut {NoStop}%
\bibitem [{\citenamefont {Beck}\ \emph {et~al.}(2007)\citenamefont {Beck},
  \citenamefont {Goncharov}, \citenamefont {Struzhkin}, \citenamefont
  {Militzer}, \citenamefont {Mao},\ and\ \citenamefont {Hemley}}]{Beck2007}%
  \BibitemOpen
  \bibfield  {author} {\bibinfo {author} {\bibfnamefont {P.}~\bibnamefont
  {Beck}}, \bibinfo {author} {\bibfnamefont {A.~F.}\ \bibnamefont {Goncharov}},
  \bibinfo {author} {\bibfnamefont {V.~V.}\ \bibnamefont {Struzhkin}}, \bibinfo
  {author} {\bibfnamefont {B.}~\bibnamefont {Militzer}}, \bibinfo {author}
  {\bibfnamefont {H.-k.}\ \bibnamefont {Mao}}, \ and\ \bibinfo {author}
  {\bibfnamefont {R.~J.}\ \bibnamefont {Hemley}},\ }\href {\doibase
  10.1063/1.2799243} {\bibfield  {journal} {\bibinfo  {journal} {App. Phys.
  Lett.}\ }\textbf {\bibinfo {volume} {91}},\ \bibinfo {pages} {181914}
  (\bibinfo {year} {2007})}\BibitemShut {NoStop}%
\bibitem [{\citenamefont {Imada}\ \emph {et~al.}(2014)\citenamefont {Imada},
  \citenamefont {Ohta}, \citenamefont {Yagi}, \citenamefont {Hirose},
  \citenamefont {Yoshida},\ and\ \citenamefont {Nagahara}}]{Imada2014}%
  \BibitemOpen
  \bibfield  {author} {\bibinfo {author} {\bibfnamefont {S.}~\bibnamefont
  {Imada}}, \bibinfo {author} {\bibfnamefont {K.}~\bibnamefont {Ohta}},
  \bibinfo {author} {\bibfnamefont {T.}~\bibnamefont {Yagi}}, \bibinfo {author}
  {\bibfnamefont {K.}~\bibnamefont {Hirose}}, \bibinfo {author} {\bibfnamefont
  {H.}~\bibnamefont {Yoshida}}, \ and\ \bibinfo {author} {\bibfnamefont
  {H.}~\bibnamefont {Nagahara}},\ }\href {\doibase
  10.1002/2014GL060423.Received} {\bibfield  {journal} {\bibinfo  {journal}
  {Geophys. Res. Lett.}\ }\textbf {\bibinfo {volume} {41}},\ \bibinfo {pages}
  {4542} (\bibinfo {year} {2014})}\BibitemShut {NoStop}%
\bibitem [{\citenamefont {Birge}, \citenamefont {Dixon},\ and\ \citenamefont
  {Menon}(1997)}]{Birge1997}%
  \BibitemOpen
  \bibfield  {author} {\bibinfo {author} {\bibfnamefont {N.}~\bibnamefont
  {Birge}}, \bibinfo {author} {\bibfnamefont {P.~K.}\ \bibnamefont {Dixon}}, \
  and\ \bibinfo {author} {\bibfnamefont {N.}~\bibnamefont {Menon}},\
  }\href@noop {} {\bibfield  {journal} {\bibinfo  {journal} {Thermochim Acta}\
  }\textbf {\bibinfo {volume} {304/305}},\ \bibinfo {pages} {51} (\bibinfo
  {year} {1997})}\BibitemShut {NoStop}%
\bibitem [{Note3()}]{Note3}%
  \BibitemOpen
  \bibinfo {note} {The only change due to addition of a constant background
  heating source to a sinusoid (e.g. $Q \propto 1+\protect \qopname \relax
  o{sin}(2\pi f t)$ or $Q \propto 100 + \protect \qopname \relax o{sin}(2\pi f
  t)$) is in time-averaged temperature profile. The dynamic temperature
  response is identical.}\BibitemShut {Stop}%
\bibitem [{\citenamefont {Kiefer}\ and\ \citenamefont
  {Duffy}(2005)}]{kiefer2005}%
  \BibitemOpen
  \bibfield  {author} {\bibinfo {author} {\bibfnamefont {B.}~\bibnamefont
  {Kiefer}}\ and\ \bibinfo {author} {\bibfnamefont {T.~S.}\ \bibnamefont
  {Duffy}},\ }\href {\doibase 10.1063/1.1906292} {\bibfield  {journal}
  {\bibinfo  {journal} {J. Appl. Phys.}\ }\textbf {\bibinfo {volume} {97}},\
  \bibinfo {pages} {114902} (\bibinfo {year} {2005})}\BibitemShut {NoStop}%
\bibitem [{Note4()}]{Note4}%
  \BibitemOpen
  \bibinfo {note} {In the case of Joule-heating, we will redefine $\omega $ as
  the angular frequency of current or voltage oscillations, which are 2-fold
  smaller than the frequency of power oscillations, meaning ``$\omega $''s in
  Eq. (\ref {eqn:C_meas_simple}) will be replaced by ``$2\omega $''s in this
  paper and in Part II.}\BibitemShut {Stop}%
\bibitem [{\citenamefont {El-Sharkawy}\ \emph {et~al.}(1984)\citenamefont
  {El-Sharkawy}, \citenamefont {Rashed}, \citenamefont {Zaghloul},\ and\
  \citenamefont {Ghoniem}}]{El-Sharkawy1984}%
  \BibitemOpen
  \bibfield  {author} {\bibinfo {author} {\bibfnamefont {A.~A.}\ \bibnamefont
  {El-Sharkawy}}, \bibinfo {author} {\bibfnamefont {I.~H.}\ \bibnamefont
  {Rashed}}, \bibinfo {author} {\bibfnamefont {M.~S.}\ \bibnamefont
  {Zaghloul}}, \ and\ \bibinfo {author} {\bibfnamefont {M.~H.}\ \bibnamefont
  {Ghoniem}},\ }\href@noop {} {\bibfield  {journal} {\bibinfo  {journal} {Phys.
  Stat. Sol.}\ }\textbf {\bibinfo {volume} {85}},\ \bibinfo {pages} {429}
  (\bibinfo {year} {1984})}\BibitemShut {NoStop}%
\bibitem [{\citenamefont {Gomi}\ \emph {et~al.}(2013)\citenamefont {Gomi},
  \citenamefont {Ohta}, \citenamefont {Hirose}, \citenamefont {Labrosse},
  \citenamefont {Caracas}, \citenamefont {Verstraete},\ and\ \citenamefont
  {Hernlund}}]{Gomi2013}%
  \BibitemOpen
  \bibfield  {author} {\bibinfo {author} {\bibfnamefont {H.}~\bibnamefont
  {Gomi}}, \bibinfo {author} {\bibfnamefont {K.}~\bibnamefont {Ohta}}, \bibinfo
  {author} {\bibfnamefont {K.}~\bibnamefont {Hirose}}, \bibinfo {author}
  {\bibfnamefont {S.}~\bibnamefont {Labrosse}}, \bibinfo {author}
  {\bibfnamefont {R.}~\bibnamefont {Caracas}}, \bibinfo {author} {\bibfnamefont
  {M.~J.}\ \bibnamefont {Verstraete}}, \ and\ \bibinfo {author} {\bibfnamefont
  {J.~W.}\ \bibnamefont {Hernlund}},\ }\href {\doibase
  10.1016/j.pepi.2013.07.010} {\bibfield  {journal} {\bibinfo  {journal}
  {Physics of the Earth and Planetary Interiors}\ }\textbf {\bibinfo {volume}
  {224}},\ \bibinfo {pages} {88} (\bibinfo {year} {2013})}\BibitemShut
  {NoStop}%
\bibitem [{\citenamefont {Ubbelohde}(1978)}]{Ubbelohde1978}%
  \BibitemOpen
  \bibfield  {author} {\bibinfo {author} {\bibfnamefont {A.~R.}\ \bibnamefont
  {Ubbelohde}},\ }\href@noop {} {\emph {\bibinfo {title} {{The Molten State of
  Matter}}}}\ (\bibinfo  {publisher} {John Wiley},\ \bibinfo {address}
  {Chichester},\ \bibinfo {year} {1978})\ p.\ \bibinfo {pages}
  {454}\BibitemShut {NoStop}%
\bibitem [{Note5()}]{Note5}%
  \BibitemOpen
  \bibinfo {note} {This is Eq. (2.3c) of Ref. \protect \rev@citealpnum
  {Kraftmakher2004}, Eq. (1) of Ref. \protect \rev@citealpnum {Bouquet2000} and
  Eq. (11) of Ref. \protect \rev@citealpnum {Sullivan1968} with $\tau _1 =
  C/K_b$, $\tau _2 \to 0$, and $K_b/K_s \to 0$}\BibitemShut {NoStop}%
\end{thebibliography}%

\end{document}